\documentstyle{report}
\catcode`\@=11
\def\euf@@ #1#2#3#4#5\cr#6{eu#6m#5}
\def\euf@@@#1#2#3{\font\@euf\expandafter\euf@@\fontname\expandafter\the#22\cr
#1#2#3=\@euf}
\def\euf{\expandafter\ifx\the\textfont2\euf@font\else\euf@init\fi
\the\textfont\euffam\fam\euffam}

\ifx\euffam\undefined
\newfam\euffam
\fi
\def\euf@init{\expandafter\let\expandafter\euf@font\the\textfont2
\euf@@@ f\textfont\euffam
\euf@@@ f\scriptfont\euffam
\euf@@@ f\scriptscriptfont\euffam}

\euf@init

\catcode`\@=12
\def\lplain{lplain}
\def\nlplain{nlplain}
\ifx\lplain\fmtname\catcode`\@=11\fi
\ifx\nlplain\fmtname\catcode`\@=11\fi

\def\got #1{{\euf #1}}

\def\[{\relax\ifmmode\@badmath\else
 \begin{trivlist}
 \@beginparpenalty\predisplaypenalty
 \@endparpenalty\postdisplaypenalty
 \item[]\leavevmode
 \hbox to\linewidth\bgroup$ \displaystyle
 \hskip\mathindent\bgroup\fi}
\def\]{\relax\ifmmode \egroup $\hfil \egroup \end{trivlist}\else \@badmath \fi}
\def\equation{\@beginparpenalty\predisplaypenalty
 \@endparpenalty\postdisplaypenalty
\refstepcounter{equation}\trivlist \item[]\leavevmode
 \hbox to\linewidth\bgroup $ \displaystyle
\hskip\mathindent}
\def\endequation{$\hfil \displaywidth\linewidth\@eqnnum\egroup \endtrivlist}
\def\eqnarray{\stepcounter{equation}\let\@currentlabel=\theequation
\global\@eqnswtrue
\global\@eqcnt\z@\tabskip\mathindent\let\\=\@eqncr
\abovedisplayskip\topsep\ifvmode\advance\abovedisplayskip\partopsep\fi
\belowdisplayskip\abovedisplayskip
\belowdisplayshortskip\abovedisplayskip
\abovedisplayshortskip\abovedisplayskip
$$\halign to
\linewidth\bgroup\@eqnsel\hskip\@centering$\displaystyle\tabskip\z@
 {##}$&\global\@eqcnt\@ne \hskip 2\arraycolsep \hfil${##}$\hfil
 &\global\@eqcnt\tw@ \hskip 2\arraycolsep $\displaystyle{##}$\hfil
 \tabskip\@centering&\llap{##}\tabskip\z@\cr}
\def\endeqnarray{\@@eqncr\egroup
 \global\advance\c@equation\m@ne$$\global\@ignoretrue }
\newdimen\mathindent
\mathindent = \leftmargini

\newtheorem{thm}{Theorem}[section]

\newtheorem{prop}[thm]{Proposition}
\newtheorem{defi}[thm]{Definition}

\newcommand{\semi}{\ltimes}
\newcommand{\ltimes}{\rhd \!\!\! <}
\newcommand{\sfrac}[2]{\mbox{$\frac{#1}{#2}$}}
\newcommand{\rd}{{\rm d}}

\title{BRST model applied to symplectic geometry}

\author{Jaap Kalkman}

\maketitle

\begin{document}

\begin{abstract}

In this thesis a finite dimensional mathematical model
is developped for cohomological field theories. These
theories are topological quantum field theories for
which there are no classical equations of motions.
They might not have direct physical relevance, but serve
to study physical methods more abstractly (I think of the
mathematical investigation of these abstract theories as
\lq second mathematization').
The BRST
cohomology entirely depends on the geometrical input of the
configuration space and the symmetry group.
A typical example is Topological Yang Mills Theory, where the
BRST operator acts as follows on the variables $A,\psi,
\omega,\phi$.
\begin{eqnarray}
s(A)=\psi-D_A\omega & & s(\omega)=-\sfrac{1}{2}[\omega,\omega]+\phi \\
s(\psi)=[\omega,\psi]-D_A\phi & & s(\phi)=-[\omega,\phi] \nonumber
\end{eqnarray}
where $A$ is a coordinate on the space of connections ${\cal A}$,
$\omega$ and $\psi$ are ghost fields associated to the invariance
with respect to gauge transformations and all transformations,
respectively, and where $\phi$ is the ghost for ghost field.
Replacing ${\cal A}$ by a finite dimensional manifold $M$ and
replacing the group of gauge transformations by a finite
dimensional group $G$, we found that the BRST algebra above
can be written as $W({\got g}) \otimes \Omega(M)$ with differential
${\rm d}_W \otimes 1  + 1 \otimes {\rm d} + \omega^a \otimes {\cal L}_a
-\phi^b \otimes \iota_b$.

Restricting
the BRST algebra of these field theories
we thus obtained that it is
isomorphic to the Cartan model for equivariant
cohomology.
Also path integrals, representing the correlation functions
of the cohomological field theory, are analyzed in detail.
For this purpose we have extended Fourier transform to
an operator acting on differential forms on a vector space,
rather than on functions only.

We also showed how the BRST algebra of Cohomological
Field Theories arises from a Lie algebra cohomology complex,
associated to the Lie algebra of $G \ltimes {\rm Diff}(M)$, provided
so called ghosts for ghosts (a symmetric algebra) are added.
We found that the difference between the Weil model and the Cartan
model for equivariant cohomology is nothing but the choice of a
basis of Lie($G \ltimes {\rm Diff}(M)$), used to write down
the Lie algebra cohomology differential.

Finally, the model is applied to configuration spaces that
are symplectic and group actions that are Hamiltonian.
In the case of a circle action,
we find a localization formula for equivariant forms
on manifolds with a boundary. This formula is used
to obtain information on the cohomology ring of the
symplectic quotient.
As an example, the cohomology rings of
symplectic quotients of ${\bf CP}^n$ by circle actions are
computed.

\end{abstract}

\tableofcontents

\chapter*{Introduction}

In this thesis I will try to shed some light into the twilight
zone between mathematics and theoretical physics.
The risk of being in a twilight zone is that you get lost,
which in this case entails uncertainty on the kind of
scientific language  to use or even on the kind of reasoning.
The pay-off
is that it is exciting to discover unexpected things.
In my case these are the rich mathematical structures that
are hidden in (our view on) nature.

The structures that I found while studying BRST quantization
(BRST are the first letters of the inventors Becchi, Rouet,
Stora and Tyutin)
of Topological Quantum Field Theories (TQFT's) include
equivariant cohomology, super Fourier transform and
double complexes. Along with some geometry they constitute
the tool-kit for the mathematical model
I developped for certain TQFT's.
This tool-kit is presented in the first chapter and provides the
reader with the mathematical background that is necessary to
read this thesis.
Although this thesis certainly is a thesis in mathematics,
some readers might welcome a background
from physics. Therefore, the rest of this introduction is devoted to
what field theory is about and why we are interested in TQFT's.

Field theories are used to describe dynamical systems (e.g.,
systems of elementary particles). They involve a
finite dimensional manifold $M$ (which is often identified
with space-time), a set of fields (often a space of
sections of some fiber bundle over $M$) and a real valued
functional ${\cal S}$ on the set of fields, called the
{\it action}. The action needs to be local which means that it
is of the form ${\cal S}=\int_M {\cal L}$, where ${\cal L}$ is
a $C^\infty(M)$-valued functional, called the Lagrangian density.
Stationary points of the action are called the classical
solutions of the field theory. They satisfy the Euler-Lagrange
equations.

\vspace{5pt}

An example of a field theory is Yang-Mills theory. Let $M$ be
a four dimensional Riemannian manifold, $G$ a
semi-simple compact Lie group and
$P \rightarrow M$ a principal $G$-bundle. Let ${\cal A}$ denote
the affine space of connections on $P$.
This is the space of fields for Yang-Mills (YM) theory.
We shall now describe the action functional.
For any connection
$A \in {\cal A}$ the curvature $F_A$ is a two form on $P$ with
values in the Lie algebra ${\got g}$ of $G$.
Let us choose a $G$-invariant metric on $P$. Then
the YM action is the square of the length of the curvature, using
this metric and the Killing metric on ${\got g}$. In a formula
\begin{equation}
   {\cal S}_{\rm YM}(A) = \sfrac{1}{4\pi^2} \int_M \mid F_A \mid^2
\end{equation}
where the integrand (obtained using  the Killing
form on ${\got g}$ and the Riemannian metric on
$P$ at each point ) is a $G$-invariant function and hence
descends to a function on $M$.
The classical solutions of YM theory are connections that
satisfy the YM equations, which includes self dual
connections on $P$.

An important feature of this field theory is that there is an
infinite dimensional group, the group of gauge transformations
${\cal G}$,
acting on the space of fields and leaving the action
invariant. In fact, the action ${\cal S}_{\rm YM}$ should be
regarded as being defined on the quotient ${\cal A}/{\cal G}$.
At least part of
the space of classical solutions is then finite dimensional
(namely the moduli space of instantons).

\vspace{5pt}

Topological Field Theories are given by actions that are
invariant under diffeomorphism groups of some kind. The name
\lq topological' is a bit misleading because there is
often much more structure involved than only a topology.
There are two types of Topological Field Theories. The first
type consists of field theories described by an action that
is invariant under Diff($M$). The second type has a constant
action and is therefore invariant under all possible transformations
of the fields. Of both types we will give an example here in order
to show the difference between
these field theories.

\vspace{5pt}

The best known example of the first type is Chern-Simons (CS) theory.
Let $M$ now be a compact three dimensional manifold, $G$ a
semi-simple compact Lie group with Lie algebra ${\got g}$.
The space of fields ${\cal A}$ is the linear space of
${\got g}$-valued one forms on $M$. The following action
on ${\cal A}$ defines CS-theory
\begin{equation}
   {\cal S}_{\rm CS}(A) = \sfrac{1}{4\pi} \int_M
    {\rm Tr}(A \wedge {\rm d}A + \frac{1}{3} A \wedge [A,A]),
\end{equation}
where Tr denotes the Killing form on ${\got g} \times {\got g}$.
Since the integrand is a differential three form, the
action is manifestly invariant under Diff($M$), the group of
all diffeomorphisms of $M$. The classical solutions are the
flat connections on the trivial bundle $M \times G$. Of
great interest is the quantum theory associated to
this field theory. The so-called quantum observables give
rise to knot invariants that are related to the Jones- and
the HOMFLY polynomials. This is the main importance of
topological field theories: although the classical theory is
rather dull, their associated quantum field theories give
rise to interesting invariants.

\vspace{5pt}

A well known example of the second type is Topological
Yang-Mills theory (TYM). The space of fields is the same as
for ordinary YM, but the action now reads
\begin{equation}
    {\cal S}_{\rm TYM}(A)= \sfrac{1}{4\pi^2} \int_M
                       {\rm Tr}(F_A \wedge F_A)
\end{equation}
There is no metric needed to define this action. The integrand
is a four form on $P$, but it is $G$-invariant and horizontal and
therefore corresponds uniquely to a four form on $M$.
It turns out that this action is independent of $A$, it only
depends on the type of bundle $P \rightarrow M$. So,
all variations of the fields $A$ are symmetries and the
symmetry group is Diff(${\cal A}$).
However, recall that the space of fields is ${\cal A}/{\cal G}$
rather than ${\cal A}$. In order to achieve this change, one
uses the larger symmetry group ${\cal G} \semi {\rm Diff}({\cal A})$
and only considers so-called basic elements with respect to the
${\cal G}$-action.

Of course, classically, TYM is not interesting at all.
All fields are stationary points, hence there are no Euler Lagrange
equations. On the other hand, the quantum observables are
extremely interesting. Witten showed in [W1] that they
correspond to the Donaldson polynomials, which are invariants for
the differentiable structure on $M$.

\vspace{5pt}

A big difference between the two types of topological field
theories is that for the first type it is essential that the
fields are really fields, i.e., that there is an underlying
manifold $M$, whereas for the second type this is not important
at all. We could as well start with some finite dimensional
manifold $X$ (representing the space of fields)
and study quantization on $X$ in the presense of a
symmetry group ${\cal G} \semi {\rm Diff}(X)$.
So, where the type 1 theories are essentially infinite dimensional,
the type 2 theories are not, except for the fact that Diff($X$)
is infinite dimensional.
The type 2 theories were baptized \lq Cohomological Field
Theories' by Witten ([W2]).

\vspace{5pt}

In this thesis we will study the BRST quantization method
applied to these Cohomological Field Theories.
We will replace Diff($X$) by a finite dimensional group $H$,
acting transitively on $X$. This leaves us with a
completely finite dimensional model for Cohomological
Field Theories. Path integrals in this model are just
integrals over $X$ and thus can be studied in great detail.
We will use them in the last chapter to prove a localization
formula for equivariant forms on manifolds with boundary.

\vspace{5pt}

To conclude this introduction we will make some comments on
notation and on prerequisites. Knowledge of differential
geometry (bundles, connections, cohomology) and of some
algebra (graded algebras, rings, modules) is certainly
necessary to read this thesis. Familiarity with
super structures (Berezin integration, super derivations)
and quantum field theory (path integrals,
correlation functions) might be helpful.

In the sequel, the symbols $M$ and $N$ denote
finite dimensional differentiable manifolds, $Z$
is always a submanifold defined by the zeroes of some set of
smooth functions and from now on $X$ denotes reduced phase spaces.
$G$, $H$ and $S$ are Lie groups and ${\cal A}$, ${\cal B}$ and
${\cal P}$ are algebras, often with a lot of gradings.
This implies that for our model we will use $G$ and $M$ rather
than the symbols ${\cal G}$ and $X$  of above (the reader
should not confuse this with the $G$ and $M$ appearing in the
type 1 theories!). Furthermore, Lie algebras are always
denoted by the gothic symbols ${\got g}$, ${\got h}$ and
${\got s}$ and elements of these Lie algebras can be traced
by searching for the greek symbol $\xi$.
The reader be warned that whenever a choice of a basis of
a vector space is involved, I will make use of the summation
convention, saying that a summation is understood wherever
the same indices appear, one as a subscript and one as a superscript.
Tensor products will always be graded and will
be taken over the real numbers ${\bf R}$
as is also the case for all vector spaces and algebras.

\chapter{The tool-kit}

The aim of this chapter is to provide the
mathematical tools used in the next
chapters to build the finite dimensional model for Cohomological
Field Theories.
The theorems 1.2.1 and 1.3.3 were published in [Ka1].

\section{Geometry}

In this first section we will introduce the main concepts
of symplectic geometry. Working in the symplectic category
has proven to be extremely useful in the past. In this
thesis it will be used in chapter two (BRST theory is
very transparent for Hamiltonian group actions) and in
chapter four (it provides nice examples of cohomology
computations).
Also in this section, we will introduce the geometric
input needed to understand the structure of path
integrals that is investigated in chapter three.

\subsection{Symplectic geometry}

Let $M$ be a symplectic manifold with symplectic form $\sigma$.
This means that $\sigma$ is closed and that, at each $x \in M$,
$\sigma_x$ is a non-degenerate antisymmetric bilinear form
on $T_xM$. In particular, this implies that $M$ is even
dimensional and orientable, since the top degree part of
exp$(\sigma)$ is nowhere vanishing. The symplectic form can
be used to associate to any function $f \in C^\infty(M)$
a vector field $V_f$ as follows. $\sigma_x$ identifies
$T_xM$ and $T^*_xM$. By definition, $V_f(x)$ is $-$d$f_x$ under
this identification. $V_f$ is called the Hamiltonian vector field
associated to the function $f$. This makes $C^\infty(M)$ into a Poisson
algebra through
\begin{equation}
         \{f,g \} := \sigma(V_f,V_g)
\end{equation}
This bracket is antisymmetric and satisfies the
derivation property. The Jacobi identity follows from
d$\sigma=0$.

Let $G$ be a Lie group acting on a symplectic manifold $M$ with
symplectic form $\sigma$. The $G$-action is called Hamiltonian
if there exists an equivariant mapping $\mu:M \rightarrow {\got g^*}$,
called the momentum mapping, such that for each $\xi \in {\got g}$
the infinitesimal action $V_\xi$ is equal to the Hamiltonian
vector field defined by the function $f_\xi: x \mapsto
<\xi, \mu(x)>$ on $M$. Thus
\begin{equation}
      \iota_{V_\xi} \sigma = -{\rm d}f_\xi \;\hspace{10pt}
               (\xi \in {\got g})
\end{equation}
where the lhs denotes contraction of $\sigma$ with the
vertical vector field $V_\xi$.

To a Hamiltonian $G$-action on $M$ one can associate a
symplectic quotient $X$ (also called the reduced phase space)
as follows. Suppose that $0 \in {\got g^*}$ is a regular value
of $\mu$. This implies that $Z=\mu^{-1}(0)$ is a
submanifold of $M$ and that $G$ acts locally free on $Z$
(see, e.g., [AM]). If $G$ is compact,
only finite stabilizer groups can occur, so $X=\mu^{-1}(0)/G$ is
well defined as an orbifold ([Sa]) and is called the
symplectic quotient of $M$ by $G$.

A group action is called symplectic if $\sigma$ is invariant.
This implies that d$(\iota_{V_\xi}\sigma)=0$ for any $\xi \in {\got g}$.
If, in addition, $H^1(M)=0$ and $G$ is compact
we obtain a map $\mu$ that can
be made equivariant by integration. So every compact
symplectic group action on a simply connected manifold is Hamiltonian
([AM],[Ki]).

An important example of a Hamiltonian group action is the
following. Let $M$ be any manifold (not necessarily symplectic),
let $G$ be any group action on $M$. Then $N=T^*M$ is a
symplectic manifold and the natural lift of the $G$-action
to $N$ is Hamiltonian. If $G$ acts freely on $M$, then the
symplectic quotient is isomorphic to $T^*(M/G)$.

\subsection{Submanifolds and Thom classes}

In physics, one is interested (when quantizing via path integrals)
in writing integrals over submanifolds as integrals over the
whole manifold. The manifolds in this context are infinite dimensional,
but for the sake of simplicity we will
stick to finite dimensional objects throughout this thesis.

Expressing integrals over submanifolds as integrals over
larger manifolds asks for representatives
of Poincar\'e duals. We shall explain this.
Let $M$ be a smooth oriented manifold of dimension $m>0$ and let
$Z \subset M$ be a smooth oriented
compact submanifold of dimension $m-n$.
Integration over $Z$ induces a linear map from $H^{m-n}(M)$
to $\bf{R}$
(assuming that $Z$ has no boundary), hence an element of
$H^{m-n}(M)^*$.
By Poincar\'e duality this element corresponds to an element of
$H_{\rm cpt}^n(M)$, which is by definition the Poincar\'e dual
[$\eta$] of
$Z \subset M$. More explicitly, if $\omega$ is any element of
$H^{m-n}(M)$, then

\begin{equation}
   \int_Z i^* \omega \; = \; \int_M \omega \wedge \eta
\end{equation}
where $i:Z \rightarrow M$ is the inclusion map. Obviously, the
support of $\eta$ may be shrunk into any open neighbourhood of $Z$
in $M$ (see, e.g., [BT]).

Let $\pi :{\cal V}
\rightarrow M$ be a vector bundle with fiber the $n$-dimensional
vector space $V$ and let $F:M \rightarrow {\cal V}$ be a
generic section.
If $\tau$ is a representative for the Thom
class of ${\cal V}$, then $F^* \tau$ is a representative for
the Poincar\'e dual of the zero locus of $F$ in $M$ ([BT]).
By definition, the Thom class is represented by forms that
give 1
when integrated over the fibers.
Stated otherwise, let $H_{\rm cv}^i({\cal V}) \rightarrow H^{i-n}(M)$
be integration over the fibers of classes that can be integrated
when restricted to the fibers
(normally it is assumed that the forms have compact
support in the fiber direction, but, as is pointed out in
[MQ], $L_2$ support will also do). This map is
an isomorphism, the Thom isomorphism, and its inverse
corresponds to $\pi^*$ followed by
multiplication with the Thom class $[\tau]$.

Representatives for the Thom class of a vector bundle are not
very easy to find , except for
the case of a trivial bundle. In this case,
the Thom class is just a normalized generator of $H^n(V)$.
Using an inner product on $V$ and an orientation, we get
a volume form $dz^1 \wedge \ldots \wedge dz^n$ and the
Thom class is represented by $f \; dz^1 \wedge \ldots
\wedge dz^n$, where $f$ is a function on $V$ such that
$\int_V f =1$.

We will restate this trivial result in a complicated way now,
the purpose being the generalization
to associated vector bundles $M \times_G V$ appearing in
chapter three, where we obtain the Mathai-Quillen
representative for the Thom class ([MQ]).
Let $z^i$ be linear coordinates on $V$, $b_i$ their dual coordinates
on $V^*$. Let $\psi^i={\rm d}z^i$ and $\bar{\psi}_i={\rm d}b_i$.
On the algebra $\Omega(M) \otimes \Omega(V) \otimes \Omega(V^*)$
consider the differential s$=$d$\otimes 1\otimes 1 +
1 \otimes $d$ \otimes 1 + 1 \otimes 1 \otimes \delta$,
where $\delta$ is defined by
\begin{equation} \label{42}
      \delta (b_i)=0,
      \hspace{1cm}
      \delta (\bar{\psi}_i)=-b_i
\end{equation}
Using all this notation, we have

\begin{prop}\label{tc1}
Let $F$ be a map $M \rightarrow V$
such that $F^{-1}(0)$ is
a manifold. $F$ can be regarded as
section $M \rightarrow M \times V$ of a trivial vector bundle.
Assume that this section is transversal to the zero section.
The differential form
\begin{equation}\label{reptc1}
        (2\pi)^{-n}
\int_{V^*}  e^{i \; s(z^j  \bar{\psi_j} - i\sum_j b_j
               \bar{\psi}_j)}
    =(\sqrt{\pi})^{-n}
       e^{-\sum_i (z^j)^2}
    {\rm d}z^1 \wedge \ldots \wedge {\rm d}z^n
\end{equation}
represents the Thom class of the vector bundle $M \times V$.
Its pull back by $F:M \rightarrow M \times V$
is a closed form in $\Omega(M)$ representing the Poincare dual
of the submanifold given by the equations $F=0$.
\end{prop}

\section{Equivariant cohomology}

Equivariant cohomology has been set up to compute cohomology of
quotient spaces of the form $M/G$, where $M$ is some manifold and $G$ a
connected
Lie group acting on $M$. In the case of a free and proper $G$-action,
$M/G$ is a manifold without singularities and one requires
equivariant cohomology to
coincide with the de Rham cohomology of the quotient space. We
shall define equivariant cohomology and we will
see that it fulfills this requirement.
Furthermore, we will introduce two very useful models in this
section.

\subsection{Topological definition}

Let $EG \rightarrow BG$ be the universal $G$-bundle, i.e. $EG$ is
contractible and every  principal $G$-bundle over some base space $B$ is
the pull back of the universal one by a map $B \rightarrow BG$. $BG$ is
called the classifying space of $G$-bundles.
Every Lie group has a universal bundle that is unique up to homotopy
(see, e.g., [Hu]). The standard example of a universal bundle
is the inductive limit of the Hopf fibrations $S^{2n+1}
\rightarrow {\bf CP}^n$, which is a model for the universal
$S^1$-bundle.

Let $M$ be a $G$-manifold. We define the associated
fibre bundle $M_G := EG \times_G M$ with fibres isomorphic to $M$ and
base space $BG$. The equivariant cohomology of $M$, $H^\ast_G (M)$, is
by definition the cohomology of the fibre bundle $M_G$:
\begin{equation}
H^\ast_G (M) := H^\ast (EG \times_G M)
\end{equation}
In the case of a free and proper group action,
$M_G$ can be seen as a fibre bundle over $M/G$ with fibre the
contractible space $EG$.
So we have $H^\ast (M_G) \cong H^\ast (M/G)$. On
the other hand, for $M = \{ x \}$ (the singleton set) we have
$H^\ast_G(M) = H^\ast (BG)$ and therefore the
equivariant cohomology of a point can be quite complicated (see, e.g.,
[AB]).

For compact connected groups $G$, there are two nice models for the
equivariant cohomology of
$G$-manifolds $M$. Compactness is really necessary for these
models (see [AB]). The models are called the
Weil model and the Cartan model. We shall describe these in detail
now. Some algebraic facts used in the next subsections are collected
in section 1.4.

\subsection{Weil model}

The Weil model for equivariant cohomology makes use of the
Weil algebra $W({\got g}) := S({\got g}^\ast ) \otimes \Lambda ({\got
g}^\ast )$. It has a ${\bf Z}$-grading by
giving the generators $\phi^a$ of
$S({\got g}^\ast )$ degree 2 and the generators $\omega^a$ of $\Lambda
({\got g}^\ast )$ degree 1 $(a=1,...,\dim {\got g}^\ast )$. The
$\omega^a$ are of course anti-commuting,
the $\phi^a $ commuting and both sets are dual to the same fixed basis
$( \xi_a)$ of ${\got g}$ (the Lie algebra of $G$). \vspace{10pt}

Suppose we are given a connection on some principal
$G$-bundle $P$. This gives rise to two maps, the curvature
${\got g}^\ast \rightarrow \Omega^2(P)$ and the connection
${\got g}^\ast \rightarrow \Omega^1 (P)$. These maps generate an
algebra
homomorphism, called the Chern-Weil homomorphism,
\begin{equation}\label{e2}
W({\got g}) = S({\got g}^\ast ) \otimes \Lambda ({\got g}^\ast )
\rightarrow \Omega (P)
\end{equation}
We shall make this map into a homomorphism of differential algebras by
defining the following differential on the  Weil algebra $W({\got g})$
(the $f^{a}_{bc}$ are the structure constants of
${\got g}$ with respect to the
fixed basis $( \xi_a)$):
\begin{eqnarray}\label{e3}
\rd_W\omega^a &=& - \sfrac{1}{2} f^{a}_{bc}
                 \omega^b \omega^c + \phi^a \\
\rd_W\phi^a &=& - f^{a}_{bc} \omega^b \phi^c \nonumber
\end{eqnarray}
where a summation is understood over indices appearing
both as a subscript and as a superscript.
This definition can be
extended to $W({\got g})$ using the fact that $\rd_W$
is a graded derivation of degree one (see section 1.4.1).
Because the relations above coincide with the definitions of the
curvature and Bianchi's identity, respectively, the map
(\ref{e2}) is (more or
less by definition of $\rd_W$) a homomorphism of differential algebras.
However, the
differential (\ref{e3}) does not give an interesting cohomology:
$H^\ast (W({\got g})) \cong {\bf R}$ as can be seen from a shift of
generators  $\phi^a \rightarrow \phi^a - \frac{1}{2} f^{a}_{bc} \omega^b
\omega^c$. It becomes interesting if we introduce two other derivations
on $W({\got g})$, the interior product $I_a$
(of degree -1) and the Lie derivative $L_a$ (of degree zero)
$\; (a=1,...,\dim {\got g})$:
\begin{eqnarray}\label{e4}
I_a \omega^b &=& \delta^b_a  \nonumber \\
  I_a \phi^b &=& 0 \\
 L_a &=& I_a \rd_W + \rd_W I_a =: [I_a ,\rd_W]^+ \nonumber
\end{eqnarray}
These operations are the algebraic analogues of interior
product of the connection 1-form and the infinitesimal generators of the
$G$-action, and the Lie derivative of differential forms.
The action of the $L_a$ is nothing but the natural (coadjoint) action
on $W({\got g})$.

The operators (\ref{e3}) and
(\ref{e4}) generate a ${\bf Z}_2$-graded Lie subalgebra
of the Lie superalgebra of all graded derivations.
In general, the Lie super bracket of two derivations $D_1$ and
$D_2$ is defined by
\begin{equation}\label{sbrac}
              [D_1,D_2] = D_1D_2 - (-1)^{{\rm deg}(D_1)
                    {\rm deg}(D_2)} D_2D_1
\end{equation}
In this case,
the bracket is easily calculated and is given by the following
formulas
(the upper + means anti-commutator, the upper - means
commutator of the operators).
\begin{eqnarray}\label{e5}
[\rd_W, L_a]^- = 0 & & [I_a, I_b]^+ = 0 \nonumber \\
{[\rd_W, I_a]}^+ =  L_a & & [ L_a,  L_b ]^- = f^{c}_{ab}
 L_c \\
{[\rd_W,\rd_W]}^+  = 0 & & [ L_a, I_b]^- = f^{c}_{ab} I_c \nonumber
\end{eqnarray}
As in the expressions above, we sum over indices
if they appear twice, one up and one down. One
sees immediately that the relations
(\ref{e5}) are independent of the choise of
a basis of ${\got g}$. They reflect the differential geometric situation
on $G$-manifolds (the bracket is just the commutator), so the algebra
(\ref{e5}) not only acts on
$W({\got g})$, it also acts on the algebra of
differential forms, $\Omega (M)$ and thus also on the tensor product
of these two algebras. We will distinguish operators
on different algebras  by using different notations.
Namely, $i_a$ for
the interior product of forms and $V_{\xi_a}$, the vertical vector field
generated by $\xi_a$ and
${\cal L}_a$ for the Lie derivative of forms in the
direction of $V_{\xi_a}$.

Finally, we are able to define the Weil model for equivariant
cohomology. The algebra of interest is the basic subalgebra of $W({\got
g}) \otimes \Omega (M)$, denoted by $(W({\got g}) \otimes \Omega
(M))_{\rm basic}$. It consists of elements annihilated by all the $I_a
\otimes 1 + 1 \otimes i_a$
and $L_a \otimes 1 + 1 \otimes {\cal L}_a$:
\[
(W({\got g}) \otimes \Omega (M))_{\rm basic} =
\]
\begin{equation}
                 = \left (\bigcap^{\dim
(G)}_{a=1} {\rm ker} (I_a \otimes 1 +1 \otimes i_a) \right )
\cap
 \left ( \bigcap^{\dim (G)}_{b=1} {\rm ker} (L_b \otimes 1 +1 \otimes
{\cal L}_b) \right )
\end{equation}
This subalgebra is stable under $\rd_W \otimes 1 + 1 \otimes \rd_M$
 (as follows from (\ref{e5})), so it is a
differential algebra. In the
sequel we shall omit the subscript $W$ and denote
this differential also by d.
The Chern-Weil homomorphism (\ref{e2}) with $P = EG$
reduces to an isomorphism on the level of cohomology if $G$ is compact
and connected:
\begin{equation}
H^\ast_d ((W({\got g}) \otimes \Omega (M))_{\rm basic}) \cong H^\ast_G
(M)
\end{equation}
For more details we refer to [AB].

\subsection{Cartan model}

Another model for equivariant
cohomology which is of considerable
interest is the Cartan model.

The map $\omega^a \mapsto 0$,
$W({\got g}) \otimes \Omega (M) \rightarrow
S({\got g}^\ast ) \otimes \Omega (M)$ induces an algebra isomorphism
\begin{equation}
(W({\got g}) \otimes \Omega (M))_{\rm basic} \cong (S({\got g}^\ast )
\otimes \Omega (M))^G
\end{equation}
where the upper $G$ means
the (infinitesimal) $G$-invariant subalgebra. This
isomorphism can be made into an isomorphism of differential algebras
by defining the following derivation on $(S({\got g}^\ast ) \otimes
\Omega (M))^G$:
\begin{eqnarray}\label{e9}
{\rm D} \phi^b &=& 0   \\
{\rm D} \eta &=& (1 \otimes \rd - \phi^b \otimes i_b)(\eta)
            \hspace{1cm}
                 (\eta \in \Omega (M))
\nonumber
\end{eqnarray}
This derivation squares to zero on the space of $G$-invariant
elements (using the fact that $\phi^b L_b \otimes 1$ acts as zero
on $S({\got g}^\ast)$).
Its cohomology equals the one of the Weil model. This model
for equivariant cohomology is called the Cartan model.

\vspace{10pt}

We shall describe how these models are related in more detail now,
thereby introducing a mathematical model for the BRST algebra of
topological field theories (more on this in chapter three).
Remember that on the algebra $A = W({\got g}) \otimes \Omega (M)$
we have the following differential:
\begin{eqnarray}\label{e10}
\rd\phi^a &=& -f^{a}_{bc} \omega^b \phi^c \nonumber \\
\rd\omega^a &=& -\sfrac{1}{2} f^{a}_{bc} \omega^b \omega^c + \phi^a \\
\rd\eta &= & {\rm exterior~differentiation~of} \;
\eta \in \Omega (M) \nonumber
\end{eqnarray}
where the $\phi^a$ are generators of $S({\got g}^\ast )$, $\omega^a$ of
$\Lambda ({\got g}^\ast )$ and $f^{a}_{bc}$ are structure constants of
${\got g}$, all defined with respect
to the same fixed basis of ${\got g}$.

The (unrestricted) BRST algebra $B$ of topological models
on quotient spaces is the same as above,
$B = W({\got g}) \otimes \Omega (M)$
(see, e.g., [OSvB]).
But the differential on it (the BRST
operator) differs:
\begin{equation}\label{e11}
\delta = \rd+ \omega^a {\cal L}_a - \phi^b i_b,
\end{equation}
where the operators ${\cal L}_a$ and $i_b$  act on differential
forms only and d acts on the
whole algebra. Maybe, we better write $\delta =
\rd_W \otimes 1 + 1 \otimes \rd_M + \omega^a \otimes {\cal
L}_a - \phi^b \otimes i_b$ instead  of
(\ref{e11}). The action of (\ref{e11}) on $B$
coincides, e.g., with [OSvB].
It is easy to check that its square equals zero.
We will show now that there is an algebra
automorphism of $A$ that carries (\ref{e10}) into (\ref{e11}).
\vspace{10pt}

Let $\psi : B \rightarrow A$ be the map
$\psi={\rm exp}(-\omega^a i_a) =
\prod_\alpha (1-\omega^\alpha \otimes i_\alpha )$. Note that $\psi $ is
degree preserving and that it differs in two ways from the map
introduced by Mathai and Quillen ([MQ], \S 5). In the first place
our map is an isomorphism on the whole algebra, rather than only
on the basic or $G$-invariant subalgebra. Secondly, we discriminate
between the \lq interior product' defined as an operation on forms and
on the Weil algebra.
The next theorem gives a natural setting for the
algebra homomorphisms of [MQ], \S 5.

\begin{thm}
$\psi $ is an isomorphism of differential algebras, so the diagram
\[ \begin{array}{rcl}
B & \stackrel{\psi }{\longrightarrow } & A \\
& & \\
\delta \downarrow & & \downarrow \rd \\
& & \\
B & \stackrel{\psi }{\longrightarrow } & A \end{array} \; \;\; \;
commutes. \]
\end{thm}

${\bf{Proof.}}$
$\psi^{-1} = {\rm exp} (\omega^a i_a)$, so it is clear that $\psi $ is
bijective. Furthermore, it is an algebra homomorphism because
all the factors $(1- \omega^\alpha \otimes i_\alpha)$ are
algebra homomorphisms.
This implies that $\psi^{-1} \circ \rd \circ \psi$ is a derivation
on $B$, so that it would be sufficient to  check that it equals $\delta$
on the generators of $B$. However, it is equally much work to verify
directly the equivalence between the two differentials.
In the sequel, we sum over roman but \underline{not} over
greek indices. We have:
\begin{eqnarray}
\delta \circ \omega^\alpha  i_\alpha &=& -\sfrac{1}{2} f^{\alpha }_{bc}
\omega^{b} \omega^{c} i_\alpha + \phi^\alpha i_\alpha - \omega^\alpha \delta
i_\alpha \\
\omega^\alpha i_\alpha \circ \delta & = & \omega^\alpha i_\alpha
\rd-\omega^\alpha \omega^a i_\alpha {\cal L}_a - \omega^\alpha \phi^b
i_\alpha i_b \nonumber
\end{eqnarray}
Subtracting these equations results in:
\begin{equation}\label{e13}
[\delta ,\omega^\alpha i_\alpha ] = -\sfrac{1}{2} f^{\alpha }_{bc}
\omega^b \omega^c i_\alpha + \phi^\alpha i_\alpha - \omega^\alpha {\cal
L}_\alpha - \omega^\alpha \omega^a [{\cal L}_a, i_\alpha ]
\end{equation}
from which we see that $[\delta ,\omega^\alpha i_\alpha ] =
(1+\omega^\alpha i_\alpha ) [\delta ,\omega^\alpha i_\alpha ]$. We want
to commute the extra term
(\ref{e13}) to the right of the product $\prod_\alpha
(1+ \omega^\alpha i_\alpha )$. So we compute
\begin{equation}
[[\delta , \omega^\alpha i_\alpha ], \omega^\beta i_\beta ] = [
-\omega^\alpha {\cal L}_\alpha , \omega^\beta i_\beta ] = -
f^{c}_{\alpha \beta } \omega^\alpha  \omega^\beta i_c
\end{equation}
Finally, we get:
\begin{eqnarray}
\delta \circ \psi^{-1} & = & \prod_\alpha (1+\omega^\alpha i_\alpha ) \{
\delta - \sum_\alpha \omega^\alpha {\cal L}_\alpha + \sum_\alpha
\phi^\alpha i_\alpha - \sfrac{1}{2} \sum_\alpha f^{\alpha }_{bc} \omega^b
\omega^c i_\alpha  \nonumber \\
& + & \sum_\alpha f^{c}_{a\alpha } \omega^a \omega^\alpha i_c -
\sum_{\alpha < \beta } f^{c}_{\alpha \beta } \omega^\alpha \omega^\beta
i_c \}= \\
&=& \psi^{-1} \circ d. \nonumber
\end{eqnarray}
$\Box$ \vspace{10pt}

As a corollary of this theorem, we get the following isomorphism:
\begin{equation}\label{isonr}
H^\ast_\delta (B) \cong H^\ast_d (A) \cong H^*(M)
\end{equation}
where the last isomorphism follows from the triviality of the cohomology
of $(\rd, W({\got g}))$.

To compute the image of the basic subalgebra of $A$, we need to know the
images of the operators
$I_a \otimes 1 + 1 \otimes i_a$ (these images already
appeared in [OSvB], but now we know where they come from).
It turns out that the
corresponding operator on $B$ only acts on $W({\got g})$. This follows
from:
\begin{equation}\label{e17}
(I_a \otimes 1) \prod_{\alpha } (1+ \omega^\alpha i_\alpha ) =
\prod_{\alpha } (1+ \omega^\alpha i_\alpha ) (I_a \otimes 1 + 1 \otimes
i_a)
\end{equation}
Furthermore, the operators
$L_a \otimes 1 + 1 \otimes {\cal L}_a$ commute with $\psi $, so the
$G$-action on both algebras is the same. This follows also from
(compare this with (\ref{e5})):
\begin{equation}\label{e18}
[\rd+ \omega^a {\cal L}_a - \phi^b i_b, I_c \otimes 1] = L_c
\otimes 1 + 1 \otimes {\cal L}_c
\end{equation}
We are now able to compute the equivalence of the  Weil model, induced
by $\psi $. The intersection of the kernels of $I_a \otimes 1$ restricts
$B$ to $S({\got g}^\ast )\otimes \Omega (M)$.
The kernels of $L_a \otimes 1+1 \otimes {\cal L}_a$
restrict it further to the  $G$-invariant subalgebra of
$S({\got g}^\ast )
\otimes \Omega (M)$. So the corresponding  subalgebra  of $A_{\rm
basic}$ is $(S({\got g}^\ast )\otimes \Omega (M))^G$, the algebra of the
Cartan model! Even more is true. The operator $\delta $ on this
subalgebra equals the differential
(\ref{e9}). We summarize this in the following theorem.
\begin{thm}
The $\omega$-independent, $G$-invariant elements of the
BRST algebra $B$ give the Cartan model for equivariant cohomology.
We have the following commutative diagram:
\begin{equation}\label{e19}
\begin{array}{rcl}
(B,\delta ) & \stackrel{\psi }{\longrightarrow } & (A,d) \\
& & \\
\uparrow & & \uparrow \\
& & \\
(S({\got g}^\ast ) \otimes \Omega (M))^G & \stackrel{\psi
}{\longrightarrow } & (W({\got g}) \otimes \Omega (M))_{\rm basic}
\end{array}
\end{equation}
where the vertical arrows are just inclusions.
The map $\psi $ between the two restricted algebras is
\[ \psi^{-1} \mid_{A_{\rm basic}} : \omega^a \mapsto 0 \]
\end{thm}
${\bf{Proof}}$. The commutativity
follows from the equations (\ref{e17})
and (\ref{e18}). The last remark
follows directly from the identity:
\begin{equation}
\psi^{-1} \mid_{A_{\rm basic}} \; =
\prod_\alpha (1+\omega^\alpha \otimes i_\alpha ) \mid_{A_{\rm basic}} =
\prod_\alpha (1-\omega^\alpha I_\alpha \otimes 1) \mid_{A_{\rm basic}}.
\end{equation}
$\Box$ \vspace{10pt}

We would like to point out here that
the isomorphism of the bottom line in (\ref{e19}) was also proved
by Mathai and Quillen ([MQ], \S 5).

\vspace{5pt}

A bit more generally we can use a parameter in the isomorphism $\psi $:
\begin{equation}
\psi_t = {\rm exp} (-t \omega^a i_a) \; : \; B \rightarrow A
\end{equation}
We can calculate what operator on $B$ corresponds to d. It
turns out that
\begin{equation}
\psi_{-t} \circ \rd\circ \psi_t = \exp ({\rm ad}(t\omega^a i_a))(\rd) =
\end{equation}
\[       =  \rd+
   t\omega^a \otimes {\cal L}_a - t \phi^b \otimes
    i_b + \sfrac{1}{2} t(1-t) f^{c}_{ab}
   \omega^a \omega^b  \otimes i_c.
\]
So, if we introduce $\delta_t$ as a notation for this differential, then
$\delta_0 = \rd , \delta_1 = \delta$.
Furthermore, we have
\begin{equation}
(I_a \otimes 1 + (1-t) \otimes i_a)  \prod_\alpha
(1+ t\omega^\alpha i_\alpha) = \prod_\alpha (1+ t\omega^\alpha i_\alpha)
(I_a \otimes 1 + 1 \otimes i_a)
\end{equation}
and
\begin{equation}
[\delta_t, I_a \otimes 1 + (1-t) \; 1 \otimes i_a] =
 L_a \otimes 1 + 1 \otimes {\cal L}_a
\end{equation}
So we obtain a family of Lie super algebras acting on $W({\got g})
\otimes \Omega (M)$, generated by
\begin{eqnarray}
& & I_a \otimes 1 + (1-t) \; 1 \otimes i_a  \nonumber \\
& & 1 \otimes {\cal L}_a +  L_a \otimes 1 \\
& & \delta_t \nonumber
\end{eqnarray}

\section{Fourier transform of differential forms}

Let $V$ be a  complex $n$-dimensional vector space and let
$\Lambda(V)$ be its Grassmannian algebra of dimension $2^n$.

Recall that Berezin integration on $\Lambda(V)$ is a linear map
from $\Lambda(V)$ to $\bf{C}$ that is zero on elements of degree
less than $n$ and is 1 on some fixed element
$\psi^1 \wedge \ldots \wedge \psi^n \in \Lambda^n(V)$.
It is called integration, because it has a lot of properties similar
to ordinary integration. E.g., a linear coordinate transformation
$A:V \rightarrow V$ must be compensated by a Jacobian. However,
this Jacobian is $\det^{-1}(A)$ instead of $\det(A)$. (By the way,
this is precisely the reason why integration of differential forms
can be defined independent of coordinates: the Jacobians cancel each
other!). A nice reference for this material is the book of Bryce
de Witt [dW].
We need the following (trivial) extension of this integration map.
\begin{equation}
\int d \psi : \Lambda(V^*) \otimes \Lambda(V)
              \rightarrow \Lambda(V^*)
\end{equation}
where $V^*$ is the linear dual of $V$. The tensor product is
a tensor product between ${\bf Z}_2-$graded algebras. Fourier transform
on Grassmann algebras can be defined in the following way

\begin{defi}
Let $\psi^1, \ldots ,\psi^n$ be generators of $\Lambda(V)$ of degree 1
such that $\int d\psi (\psi^1 \wedge \ldots \wedge \psi^n) = 1$ and let
$\bar{\psi}_1, \ldots , \bar{\psi}_n$ be their duals in
$\Lambda^1(V^*)$.

For every $\eta \in \Lambda(V)$, Fourier transform
${\cal F} : \Lambda(V) \rightarrow \Lambda(V^*)$ is defined by
\begin{equation}
{\cal F}(\eta) = \int d \psi \; (\eta \wedge e^{i \; \bar{\psi}_j
             \otimes \psi^j})
\end{equation}
where ${\rm exp}(i \; \bar{\psi}_j
\otimes \psi^j) \in \Lambda(V^*) \otimes
\Lambda(V) $ is given by the well known power series, which in this
case stops at the n-th power.
\end{defi}

Of course, as it stands it is just a copy of the definition
of ordinary Fourier transform. The next proposition shows
that it has also properties like ordinary Fourier transform.

\begin{prop}
If $\dim (V)$ is even, then
${\cal F}^2 : \Lambda(V) \rightarrow \Lambda(V)$ equals the
identity.
\end{prop}

${\bf Proof.}$ $\cal{F}$ is a linear map, so it suffices to check the
statement on homogeneous elements in $\Lambda(V)$. Let $
\eta =\psi^1 \wedge \ldots \wedge
\psi^k$ be an element of $\Lambda^k(V)$. The component of
$\eta \cdot {\rm exp}(i \; \bar{\psi}_j \otimes \psi^j)$ that is in
$\Lambda(V^*) \otimes \Lambda^n(V)$ is $(i)^{n-k}
(-1)^{\frac{1}{2}(n+k-1)(n-k)} \bar{\psi}_{k+1} \wedge \ldots \wedge
\bar{\psi}_n \otimes \psi^1 \wedge \ldots \wedge \psi^n$.

Thus,
\begin{equation}
{\cal F} (\psi^1 \wedge \ldots \wedge \psi^k) =
(i)^{n^2 - k^2} \bar{\psi}_{k+1} \wedge \ldots \wedge \bar{\psi}_n \in
\Lambda(V^*).
\end{equation}
Applying Fourier transform once again, we obtain
\begin{equation}
{\cal F}^2 (\psi^1 \wedge \ldots \wedge \psi^k) =
(i)^{n^2 - k^2} (i)^{k^2} \psi^1 \wedge \ldots \wedge \psi^k.
\end{equation}
For $n$ even, the prefactor equals $1$, so ${\cal F}^2$ is the identity.
$\Box$
\vspace{10pt}

{\bf Remarks}

1.1) Of course, so far we have just done linear algebra. Nevertheless,
combining this with ordinary Fourier transform, it will turn out to
 be independent of the choice of a measure and therefore very useful.

1.2) The factor $i$ in the exponent is only meant to give a nice prefactor
when computing ${\cal F}^2$. It is of no importance for the convergence
of the integral as it is in the ordinary case.

1.3) We have stressed earlier the fact that we should rather be working
in an infinite dimensional context. It should not surprise me
very much if this Fourier theory will turn out to be very useful
in that context also.
 E.g., differential forms of finite codegree can easily be
obtained from differential forms of finite degree, using Fourier
transform.
\vspace{10pt}

We will now combine this definition with ordinary Fourier transform
 to obtain Fourier transform of differential forms. We will take the
Schwartz functions, denoted by ${\cal S}(V)$, as the domain
for Fourier transform.
Differential $k-$forms, $\Omega_s^k(V)$, can be seen as elements
of ${\cal S}(V) \otimes \Lambda^k(V^*)$.
Note that, although the notations are very similar, this algebra
has little to do with the Weil algebra used in the previous
sections.
Combined Fourier transform
maps this space to ${\cal S}(V^*) \otimes \Lambda^{n-k}(V)$:
\[ {\cal F}(f \otimes \eta)(b) = \int_V
   f \; e^{i<b \mid \cdot >} \otimes \eta e^{i \omega}
\]
where $\int_V$ is integration of differential forms and
$\omega$ is the canonical symplectic 2-form on $V \times V^*$.
In coordinates $z^i$ and differentials $\psi^i
={\rm d}z^i$, $\omega =
\bar{\psi_j} \otimes \psi^j \;$, $ \langle b \mid \cdot \rangle $ means
$b_iz^i$ and integration of differential forms boils down to
ordinary integration over the $z^i$ and Berezin integration
over the $\psi^i$.

A lot of properties of Fourier transform on functions extend to this
combined Fourier transform. E.g., it is possible to extend the
definition of the convolution product such that it is the Fourier image
of the wedge product of differential forms.
The convolution of a $k-$form
and an $l-$form is a $(k+l-n)-$form.
Recall that convolution between two functions $f$ and $g$ in
${\cal S}(V)$ is defined by $f*g(y) = \int_V f(x)g(y-x)dx$. Therefore,
it is natural to define for $\eta, \zeta \in \Lambda(V)$:
\begin{equation}
(\eta * \zeta)(\phi) := \int \eta (\psi) \wedge \zeta (\phi - \psi)
d \psi
\end{equation}
where $\eta(\psi)$ means, $\eta$ expressed in terms of generators
$\psi^j$. The $\phi^i$ are just other names for the same generators,
as it is the case in the definition of ordinary convolution.
Here,
$\zeta(\phi - \psi)$ means, substitute $\phi^i - \psi^i$ wherever
a $\psi^i$ occurs in $\zeta (\psi)$.
It is not very difficult to prove that for every $\eta, \zeta \in
\Lambda(V)$ we have
\begin{equation}
{\cal F}(\eta \wedge \zeta) = {\cal F}(\eta) * {\cal F}(\zeta)
\end{equation}

Of course, we can combine this convolution product with the ordinary
one to obtain a convolution product on the algebra of differential
forms ${\cal S}(V) \otimes \Lambda(V)$. Note that the top form
$\psi^1 \wedge \ldots \wedge \psi^n$ is the unit element for the
convolution product in $\Lambda(V)$, whereas for the product on
${\cal S}(V)$ the unit is not contained in ${\cal S}(V)$ (it is
the Dirac distribution).
\vspace{10pt}

${\bf Example}$.
Suppose $ \dim(V)=4; \psi^1, \ldots ,\psi^4$ are generators of $\Lambda
(V)$ such that $\int \psi^1 \psi^2 \psi^3 \psi^4 d\psi =1$.
Then,
\[{\cal F}(\psi^1 \psi^2)= \int \psi^1 \psi^2 e^{i\bar{\psi}_j \psi^j}
d\psi =
\]
\[ =\int \psi^1 \psi^2 \; (\sfrac{-1}{2!})(\bar{\psi}_3 \psi^3 +
\bar{\psi}_4 \psi^4)^2 d\psi = \bar{\psi}_3 \bar{\psi}_4
\]
and
\[
{\cal F}(\psi^1)= -i \; \bar{\psi}_2 \bar{\psi}_3 \bar{\psi}_4 ,
\hspace{20pt}
{\cal F}(\psi^2)= i\; \bar{\psi}_1 \bar{\psi}_3 \bar{\psi}_4.
\]
Now, let us calculate the convolution product
\[ \!\! {\cal F}(\psi^1) * {\cal F}(\psi^2) = \int \phi_2 \phi_3 \phi_4
(\bar{\psi}_1 - \phi_1)(\bar{\psi}_3 - \phi_3)(\bar{\psi}_4 -\phi_4)
d\phi = \bar{\psi}_3 \bar{\psi}_4.
\]
So for this particular case we have verified that the Fourier image of
the wedge product is the (super-)convolution product.
\vspace{10pt}

Another property of Fourier transform that we would like to
extend is the following. If $z^i$ are coordinates on $V$ and $b_i$
are the dual coordinates on $V^*$
, then it is well known
that ${\cal F}(\frac{\partial f}{\partial z^j}) =
(-i  b_j) {\cal F}(f)$. Thus,
multiplying with $b_i$ is a derivation for the convolution product.
On the (super-)algebra $
\Omega_s(V)= {\cal S}(V) \otimes \Lambda(V^*)$ there exists
a (super-)derivation with square zero, namely the de
 Rham differential d.
We would like to know its Fourier image.
Let us define on ${\cal S}(V^*) \otimes \Lambda(V)$ the Koszul
differential $\delta$ as follows.
$\delta$ is the derivation of
degree $-1$ acting on generating elements by
\begin{equation}
\delta(f) = 0  \hspace{80pt}  (f \in {\cal S}(V^*))
\end{equation}
\[ \delta(f \otimes \bar{\psi}_i) = - f b_i \otimes 1
    \hspace{25pt} (i=1, \ldots ,\dim(V))
\]
Here, $\bar{\psi_i}$ is the image of $b_i$ under the inclusion
$V \rightarrow \Lambda(V)$. Obviously, the definition of
$\delta$ does not depend on the choice of linear coordinates
on $V^*$.
We will prove now that this is the Fourier image of the
de Rham differential.

\begin{thm}
${\cal F} \circ \rd= \delta \circ {\cal F}$
\end{thm}
${\bf Proof.}$

\[\delta \circ {\cal F} (f \otimes \eta) =
\delta (\int f \; e^{i\; <b \mid z>} \otimes \eta \;
e^{i\; \bar{\psi}_j \psi^j} dz \; d\psi) =
\]
\[\int f(z) e^{i\; <b \mid z>} (-ib_j) \otimes \psi^j \eta
e^{i\; \bar{\psi}_j \psi^j} dz \; d\psi =
\]
\[\int \frac{\partial f}{\partial z^j} \;
 e^{i\; <b \mid z>} \otimes \psi^j \;
\eta \; e^{i\; \bar{\psi}_j \psi^j} dz \; d\psi =
{\cal F} \circ \rd (f \otimes \eta). \Box
\]

\vspace{5pt}
{\bf Remarks}

1.4) One may wonder why $\delta$ depends on the linear
structure of $V^*$, whereas d does not ($\delta$ only
commutes with linear diffeomorphisms, d commutes with
all diffeomorphisms). This is because Fourier transform,
which carries the one into the other, depends on the
linear structure.

1.5) In the BRST model of topological theories, the Fourier transform
will be used as follows. The algebra $\Omega(V) \otimes
\Omega(V^*) $ can be given a double complex structure
using the differentials $\rd \otimes 1$ and
$(-1)^p \otimes \delta$ ($p$ being the degree operator). The
sum $s$ of these two differentials is part of the BRST operator.
Using $s$, we can define (extended) Fourier transform
$\bar{{\cal F}} : \Omega_s(V) \otimes \Omega_s(V^*)
\rightarrow \Omega_s(V)$ by integration over $\Omega_s(V^*)$,
after multiplication with exp$(i  s(z^j \bar{\psi}_j))$.
 From the proposition above, it follows that
$\bar{{\cal F}} \circ s = \rd \circ \bar{{\cal F}}$.
BRST theory uses the map $\bar{{\cal F}}$ to obtain
d-closed differential forms from rather simple
$s$-closed expressions.

\section{Some algebra}

In this section we recall some elementary facts on
superstructures and double complexes
that will be used throughout this thesis.

\subsection{Superstuff}

A super vector space is a vector space with a ${\bf Z}_2$-grading.
A superalgebra is a super vector space such that the product
respects the grading. In this section we will consider
superalgebras of type ${\cal A}=\Lambda V \otimes SW$, where
$V$ and $W$ are vector spaces.
${\cal A}={\cal A}^+ \oplus {\cal A}^-$, ${\cal A}^+$ being
the even part $\Lambda^{\rm even}V \otimes SW$, ${\cal A^-}$ being the
odd part $\Lambda^{\rm odd}V \otimes SW$.
The supercommutator on a
superalgebra is defined by (\ref{sbrac}). If it vanishes for
any two elements of the algebra, the superalgebra is called
supercommutative. Our algebras ${\cal A}$ are supercommutative.
Another example of a supercommutative
superalgebra is the space of differential forms on a
manifold.

Each linear map d  on ${\cal A}$
can be written as (d$^0_0$,d$^0_1$,d$^1_0$,d$^1_1$),
where d$^1_0:{\cal A}^+ \rightarrow {\cal A}^-$, etc.
A linear map d is called a derivation if all d$_i^j$ satisfy
\begin{equation}
   {\rm d}_i^j(ab) = ({\rm d}_i^ja) b +(-1)^{(i+j){\rm deg}a}a
                      ({\rm d}_i^jb)
\end{equation}
for homogeneous (i.e., either even or odd) elements $a \in {\cal A}$.
Let us write d$_+={\rm d}^0_0 +{\rm d}_1^1$ and d$_-={\rm d}_0^1 +
{\rm d}_1^0$. Then d$=$d$_++$d$_-$. This endows the space of
linear maps End(${\cal A}$), and thus also the subspace of
derivations Der(${\cal A}$), with a superstructure. This will be
denoted by Der$({\cal A}) =
{\rm Der}_-({\cal A}) \oplus $Der$_+({\cal A})$.
The following statements are easy to prove

\begin{prop}

a) The supercommutator of two derivations is again a derivation.
Hence Der(${\cal A}$) has a Lie superalgebra structure.

b) Any linear map $V \oplus W \rightarrow {\cal A}$ can be extended
uniquely to a derivation on ${\cal A}$. In particular, a derivation
is already defined through  its action on $\Lambda^1V \otimes S^1W$.
\end{prop}

\subsection{Double complex structures and applications}

In this section we would like to show that for double
complexes, satisfying certain conditions, the cohomology
is given by the $E_2$-term of the spectral sequence.
However, we shall not use the theory of spectral sequences,
but prove all statements \lq by hand'.
Furthermore, we will give some useful examples.

Let ${\cal A}=\oplus_{p,q} {\cal A}^{p,q}$ be an algebra with two
non-negative gradings ($p,q \in {\bf Z}_{\geq 0})$.
Let d$_1$ and d$_2$ be two differentials, i.e., d$_i^2=0$
and both are derivations,
\[   {\rm d}_1:{\cal A}^{p,q} \rightarrow
     {\cal A}^{p+1,q},
      \hspace{1cm}
   {\rm d}_2: {\cal A}^{p,q} \rightarrow {\cal A}^{p,q+1}
\]
Furthermore, let us assume that d$_1$d$_2 =-$d$_2$d$_1$.
Then D$=$d$_1+$d$_2$ is again a differential, increasing the total
degree $p+q$ by $1$.
Thus we obtain three different cohomologies on ${\cal A}$,
two bi-graded ones ($H_{d_1}^{\bullet ,q}({\cal A})$ and
$H_{d_2}^{p,\bullet}({\cal A})$) and the single graded total
cohomology $H_D^\bullet ({\cal A})$.

We will need the following filtration on
$H_D({\cal A})$:
for all $P \geq 0$, define
\[ {\cal A}_P^r=\left ( \bigoplus_{p+q=r} {\cal A}^{p,q} \right)
        \bigcap \left( \bigoplus_{p \leq P} (\oplus_{q}
                 {\cal A}^{p,q}) \right )
\]
We will fix $r \geq 0$ and often omit
the superscript of ${\cal A}_P^r$ in
the sequel. The inclusions
$H_D({\cal A}_P) \subset H_D({\cal A}_{P+1})$ induce a filtration
on $H^r_D({\cal A})$. Let $GH_D^r({\cal A})$ denote the associated
graded algebra.

\begin{thm}\label{spseq1}
There exists a natural map
\begin{equation}\label{dcmap}
      H_D({\cal A}^r_P) \rightarrow H_{d_2}^{P,r-P}(H_{d_1}({\cal A}))
\end{equation}
Furthermore, if, whenever $p-q$ is odd, $H^{p,q}_{d_1}({\cal A})=0$,
then this map is surjective and we have an isomorphism of
graded vector spaces
\begin{equation}
   GH_D^r({\cal A}) \simeq H_{d_2}(H_{d_1}({\cal A}^r)),
\end{equation}
where ${\cal A}^r= \oplus_{p+q=r} {\cal A}^{p,q}$.
\end{thm}

{\bf Proof.}
Suppose $\eta = \sum_{p+q=r} \eta^{p,q} \in {\cal A}_P$
satisfies D$\eta=0$. Thus, d$_1(\eta^{p,q})=-{\rm d}_2(\eta^{p+1,q-1})$.
Let $Q=r-P$. We will prove that $\eta \mapsto \eta^{P,Q}$ induces
a map between cohomologies.

As d$_1(\eta^{P,Q})=0$ and d$_2(\eta^{P,Q}) = {\rm d}_1(-
\eta^{P-1,Q+1})$, $\eta^{P,Q}$ represents a class in
$H_{d_2}(H_{d_1}({\cal A}))$, indeed. Furthermore, it
is easily checked that $\eta$ and $\eta + {\rm D}\rho$ are
mapped to the same class. This proves the first part of the
theorem.

Next, suppose $H_{d_1}({\cal A}^{p,q})=0$, whenever $p-q$ is odd.
Let $\eta^{P,Q}$ represent a non-zero class
in the rhs of (\ref{dcmap}), i.e., d$_1 (\eta^{P,Q})=0$ and
d$_2(\eta^{P,Q})={\rm d}_1(\eta^{P-1,Q+1})$ for some
$\eta^{P-1,Q+1} \in {\cal A}$. We may assume that
$P-Q$ is even (otherwise, [$\eta^{P,Q}]=0$).
 From
\[       {\rm d}_1({\rm d}_2(\eta^{P-1,Q+1}))
             =-{\rm d}_2({\rm d}_1(\eta^{P-1,Q+1}))
        =-{\rm d}_2^2(\eta^{P,Q})=0
\]
we get the existence of
$\eta^{P-2,Q+2}$ such that ${\rm d}_2(\eta^{P-1,Q+1}) =
{\rm d}_1(\eta^{P-2,Q+2})$. Again ${\rm d}_1({\rm d}_2(
\eta^{P-2,Q+2})) = 0$ and we find an $\eta^{P-3,Q+3}$ etc.
Putting things together, we obtain $\eta= \sum_i^P (-1)^i
\eta^{P-i,Q+i}$, which is, by construction, closed under D.

The last statement in the theorem follows directly from
the fact that the kernel of (\ref{dcmap}) is given by the
inclusion $H_D({\cal A}_{P-1}^r) \rightarrow H_D({\cal A}_P^r)$.
$\Box$

\vspace{5pt}

{\bf Remarks}

1.6) Obviously, $\eta$ is far from being unique. Its image in
$GH_D({\cal A})$, however, is unique. The method of finding
a representative $\eta$ is called the zig-zag construction ([BT]).

1.7) Without any restrictions on the cohomology of d$_1$,
$GH_D({\cal A})$ can be computed using a spectral sequence $E_i$.
Fortunately, we will not need this full theory. However,
sometimes we will use terminology like: \lq the spectral sequence
degenerates at $E_2$'. The meaning of these words is just in
the theorem. By the way, using spectral sequences gives an
easy proof of the theorems in this section.

1.8) Of course, if we substitute \lq $p-q$ is odd' by \lq $p-q$
is even', we obtain a similar theorem.

\vspace{5pt}

{\bf Example}. As an example for later use, consider the Cartan model
$(S({\got g^*}) \otimes \Omega(V))^G$
for a vector space $V$, on which a compact connected group $G$
acts linearly. On $V$ we allow only differential forms that are
square integrable. This implies that $H^i(V)=0$, except for
$i=n=$dim$(V)$.
We introduce the following two gradings. The first one is the sum
of the
polynomial degree of $S({\got g^*})$ and the
form degree on $V$. The second one is only the polynomial degree.
Note that the sum of the two degrees equals the degree defined
earlier and that the two components of the Cartan differential,
d and $\phi^b \otimes \iota_b$, are
differentials of degree $(1,0)$ and $(0,1)$, respectively.
Applying the theorem in this case, and using the fact that
for compact groups each de Rham cohomology class contains a
$G$-invariant representative gives that the equivariant
cohomology of $V$ is (as vector space) $S({\got g^*})^G
\otimes H^n(V)$.

We will use this in chapter three as follows. The
differential form (\ref{reptc1}) represents a generator of
$H^n(V)$. Formally, it can therefore be extended, using the zigzag
construction, to an equivariantly closed differential form.
In chapter three we will solve this zigzag problem using
our BRST model and (super) Fourier transform. It turns out
that the representative is the same as the one in [MQ].

\vspace{5pt}

By using similar arguments as in the proof above, we obtain
\begin{thm}\label{dc2}
Suppose $H^{p,q}({\cal A})=0$ for all $p \neq P$. Then
\begin{equation}
   GH_D({\cal A}) \simeq H_{d_2}^{r-P}(H^P_{d_1}({\cal A}))
\end{equation}
is an isomorphism of graded vector spaces.
\end{thm}

\vspace{5pt}

{\bf Example}. Remember from section 1.2 the Weil algebra
$W({\got g})=S({\got g^*}) \otimes \Lambda({\got g^*})$,
generated by $\phi^a$ and $\omega^a$, and equipped
with the differential d$_W$.
Let us introduce the degree $p$ as the polynomial degree
of $S({\got g^*})$ and $q$ as the sum of $p$ and the
exterior degree on $\Lambda({\got g^*})$. The total degree
$p+q$ agrees with the degree defined earlier on $W$.
We will define two anti-commuting differentials that add up to d$_W$,
thereby giving the Weil algebra a double complex structure.
\begin{eqnarray}\label{d1d2}
   {\rm d}_1 \omega^a = \phi^a & & {\rm d}_2 \omega^a =
                       -\sfrac{1}{2} f^a_{bc} \omega^b \omega^c \\
   {\rm d}_1 \phi^a = 0 & & {\rm d}_2 \phi^a = -f^a_{bc}
                       \omega^b \phi^c \nonumber
\end{eqnarray}
The second one is the Lie algebra cohomology differential with
values in the representation $S({\got g^*})$.
Since $H_{d_1}(W)={\bf R}$, we can apply the theorem, obtaining
$H_{d_W}(W)={\bf R}$. We found this result earlier by making
a shift in the generators.

To look at the Weil differential algebra in this way is very
important in understanding how we arrive, in chapter three, from
Lie algebra cohomology to equivariant cohomology by
adding 'ghosts for ghosts' (the $\phi^a$) and
restricting the algebra ($\omega$-independent, $G$-invariant
elements only).

\chapter{BRST theory}

In this chapter the foundations of BRST theory will be
discussed and we will set the stage for our model to
be described in the next chapter.
Although we will follow in our description of the BRST
algebra general concepts on how to define these algebras,
our definition involves non-trivial vector bundles and is therefore
more geometric. The results of
sections 2.2 and 2.3 as well as those of sections 3.1 and 3.2
of the next chapter can be found in [CK].

\section{Historical background}

The roots of BRST theory date from the late sixties, when
physicists added anti-commuting variables (called
Faddeev-Popov ghosts) to the classical action in order to
obtain a quantum action that gave rise to convergent path
integrals. This procedure (which is part of the path integral
quantization) can be explained in a simple finite dimensional
situation, which already features the BRST symmetry.

Let $G$ be a Lie group acting isometrically on a
Riemannian manifold $M$.
For simplicity, assume that the action is free and that
$\pi:M \rightarrow B$ is the associated principal fibration.
Suppose $U \subset B$ is an open subset such that
$\pi^{-1}(U) \simeq S_U \times G$, $S_U \subset M$ being
some (slice) submanifold given by the
zeroes of the functions $g^i$ $(i=1, \ldots,
{\rm dim}(G))$. We would like to integrate $G$-invariant
functions of the form $e^{i{\cal S}}$ over $\pi^{-1}(U)$.
Let $(\xi_j)$ be an orthonormal basis of ${\got g}$
with respect to some inner product,
let $f^i_{jk}$ denote the structure constants
and let $X_j=V_{\xi_j}$ denote the
generating vector fields of the $G$-action. Then
\begin{equation}
   \int_{\pi^{-1}(U)} e^{i{\cal S}} = {\rm vol}(G) \int_{\pi^{-1}(U)}
      \delta(g^i) \; {\rm det}(X_j (g^i)) e^{i{\cal S}},
\end{equation}
where $\delta(g^i)$ is the pull back by $g^i:\pi^{-1}(U)
\rightarrow {\bf R}$ of the Dirac delta function (readers
that do not like to regard this as a function may consider to
skip this introductory model).
The equation can be derived using local coordinates in a
straightforward manner.
Note that the rhs is independent of the choice of $g^i$.
In this simple model, $M$ represents the space of paths,
$g^i$ are called gauge functions and ${\cal S}$ is the classical
action. The obvious advantage of the rhs expression is that
we can ommit the vol$(G)$ factor (which is infinite in field
theory), thereby removing at least one obstacle for a finite answer.

Next, we will write $\delta(g^i)  {\rm det}(X_j (g^i))$ as an
exponential, using three new types of variables, $b_i$,
$\bar{c}_i$ and $c^i$ $(i=1, \ldots ,{\rm dim}(G))$.
The $b_i$ are commuting variables, the $\bar{c}_i$ and $c^i$
are anti-commuting. Using Berezin integration, we obtain
\begin{equation}
   \int_{\pi^{-1}(U)} e^{i{\cal S}} = {\rm vol}(G) \int_{\pi^{-1}(U)}
     e^{i({\cal S}+b_k g^k + \bar{c}_i X_j (g^i) c^j)}
\end{equation}
The exponential on the rhs is called the quantum action,
the $c^i$ are Faddeev-Popov ghosts and the $\bar{c}_i$ are
anti-ghosts.

We will now show where the BRST symmetry is in this simple model.
Let us define the following differential
\begin{eqnarray}
   s(f) = X_j(f) \; c^j & & s(\bar{c}_i)=-b_i \\
   s(c^k)=-\sfrac{1}{2} f^k_{ij} c^i c^j & & s(b_i)=0  \nonumber
\end{eqnarray}
It is easy to check that $s^2=0$, that $s({\cal S})=0$ and
that the quantum
action is nothing but ${\cal S}+s(-\bar{c}_i g^i)$.
To say it in other words: ${\cal S}$ defines a cohomology class for $s$
and the integral remains unchanged if we take another
representative. The cohomology involved is called the BRST
cohomology and $s$ is the BRST operator.

It is precisely this infinitesimal symmetry that was
discovered in 1975 by Becchi, Rouet and Stora ([BRS]) in the quantum
action of certain field theories. Independently, Tyutin wrote
an unpublished preprint on the same, hence the name BRST.
During the late seventies it were mainly Russian physicists
developing BRST related theories ([BV],[FF]). It soon became
clear that BRST theory had a symplectic
counterpart, that it could explain the multi ghost
interactions of, e.g., supergravity and that BRST theory
was important to obtain quantum actions  and to prove
renormizability.

New impetus came from Marc Henneaux' thesis
published in [He], where
he described a BRST theory for Hamiltonian systems with
constraints. It turned out that, in the case of first class
constraints, BRST theory involves the use of superPoisson
structures and supercanonical transformations rather than ordinary
Poisson algebras.
After this work of Henneaux, BRST became an industry.
Mathematicians were involved to develop the
Poisson algebraic and the differential algebraic parts, while
in the physics community BRST theory became an important
quantization concept. Part of the history is captured
in [St].

\section{BRST complex for group symmetries}

The aim of this section is to describe BRST theory associated
to a symplectic manifold $N$, called the phase space, and a
set of first class constraints.
Since we will be interested in constraints coming from group
actions, we will deal mostly with the group case, making only some
remarks on the general situation.

First class constraints are just functions on $N$
satisfying certain conditions. The zero set $Z$ of these functions
is called the constraint manifold (provided that it is a
manifold). In the sequel we will write
constraints as a short hand for first class constraints.
BRST theory assigns to a given set of constraints a
BRST charge $Q$. This charge $Q$ is an element of a super
Poisson algebra extension ${\cal P}$ of $C^\infty(N)$. $Q$ is a
natural and geometrical object in the sense that if the constraint
manifold is represented by another set of constraints,
then the two BRST charges are related by a supercanonical
transformation.

Let $(N,\sigma)$ be a symplectic manifold. Then $C^\infty(N)$
is a Poisson algebra with Poisson bracket $\{ \cdot,\cdot\}$.
A collection of functions $(f_a)$ is called
a set of first class constraints if (remember we use the
summation convention)
\begin{equation}\label{struc}
\{ f_a, f_b \} = c_{ab}^c f_c
\end{equation}
for certain $c_{ab}^c \in C^\infty(N)$. For
simplicity we will assume that $N$ is finite dimensional and
that $f_a$ is a finite collection of functions.
Condition (\ref{struc}) implies that the ideal $I \subset C^\infty(N)$
generated by the constraints is a Poisson subalgebra.

Constraints are called regular or independent if $Z$ is
a manifold
of codimension equal to the number of constraints.
They are called weakly regular or reducible if $Z$ is
a manifold of codimension smaller than the number of constraints.
The latter case will be the subject of the next section.

Let $S$ be a Lie group acting Hamiltonially on $N$.
The components of the associated momentum map,
well defined after introducing a basis $(\xi_a)$ of
${\got s}=$Lie$(S)$, satisfy (\ref{struc}).
The $c_{ab}^c$ are constants in this case
(the group case) and equal the structure constants
$f^c_{ab} $ of the Lie algebra
${\got s}$, using the same base.
We will assume in this section
that the constraints are regular, i.e., 0
is a regular value of the momentum map.

By definition, the BRST algebra ${\cal P}$ in the group case is
$\Lambda {\got s} \otimes
\Lambda {\got s^*} \otimes C^\infty (N)$.
This algebra has two gradings, induced by the degrees of
the Grassmann algebra elements. They are called the anti-ghost
number and ghost number, respectively. The basis elements
$\xi_a$, regarded as members of $\Lambda^1 {\got s}$, have
anti-ghost number $1$. The elements $\omega^a$ of the
dual base are in $\Lambda^1 {\got s^*}$ and have ghost number $1$.

In [KS] the following two differentials are defined on ${\cal P}$.
The first one, $\delta$, is zero on $\Lambda {\got s^*}
\otimes C^\infty(N)$ and maps $\xi_a \mapsto f_a$.
It lowers the anti-ghost degree by $1$.
$\delta$ is called the Koszul operator and the complex is called
Koszul resolution.
The second differential, d,
is the Lie algebra cohomology operator with values in
the representation $\Lambda {\got s} \otimes C^\infty (N)$.
It has ghost degree $1$ and acts on generators as
\begin{eqnarray}
{\rm d} \; (\xi_a \otimes 1 \otimes f)& = &
  f_{ab}^c \xi_c \otimes \omega^b \otimes f
  -\xi_a \otimes \omega^b \otimes \{f_b,f\} \\ \nonumber
{\rm d} \; (1 \otimes \omega^c \otimes 1)& = & -1 \otimes
                \sfrac{1}{2}
  f_{ab}^c \omega^a \wedge \omega^b \otimes 1
\end{eqnarray}

One can check that $\delta^2={\rm d}^2=0$ and that d$\delta =
-\delta$d. Thus the derivation $D=$d$+\delta$ is also a
differential. $D$ is the BRST operator.
It has total degree 1 if we define the total degree to be
the difference of the ghost and anti-ghost degrees.
Obviously, $({\cal P},D)$ has a double complex structure, so its
cohomology
$H_D({\cal P})$ can be computed using a spectral sequence.
It turns out that, because of the regularity condition,
$\delta$ has only non-zero cohomology in anti-ghost degree zero and
that it equals $\Lambda{\got s^*} \otimes C^\infty(N)/I$.
Therefore, the spectral sequence degenerates at $E_2$. Thus
\begin{equation}
   H_D({\cal P})=H_{\rm d}(H_\delta({\cal P}))=
    H_{\rm d}(\Lambda{\got s}^* \otimes C^\infty(N)/I)
\end{equation}
Using $C^\infty(N)/I \simeq C^\infty(Z)$, we obtain for the
BRST cohomology in degree zero $H^0_D({\cal P}) \simeq
C^\infty(Z/S)$, the algebra of functions on the reduced phase space.

\vspace{5pt}

We will now give a
nice formulation of the differential $D$
in terms of
super Poisson algebras.
A super Poisson algebra is an algebra with a
{\bf Z}$_2$-grading and a super Lie algebra structure that
is compatible with the ring structure in the sense that a
graded Leibnitz rule holds:
\begin{equation}
\{a_1\cdot a_2,a_3\}=a_1\cdot\{a_2,a_3\}+(-1)^{{\rm deg}(a_2)
                         \cdot {\rm deg}(a_1)}
                    a_2\cdot\{a_1,a_3\}
\end{equation}
For odd elements the Poisson bracket is commutative,
for any other pair of homogeneous elements the Poisson bracket
is anti-commutative. Let us introduce a super Poisson structure
on ${\cal P}$ that extends the one on $C^\infty(N)$. Besides the
bracket between two functions the only non-vanishing bracket
between generators is $\{\xi_a,\omega^b\}=\{\omega^b,
\xi_a\}=\delta_a^b$.
This determines the Poisson bracket completely
using the Leibnitz rule.
By definition, the BRST charge is a $Q \in S$ such that $D=\{Q,
\cdot \; \}$. As a little miracle, it exists and equals
$Q=f_a \omega^a - \frac{1}{2} f_{ab}^c
\xi_c \omega^a \omega^b$.
It has total degree 1
and satisfies $\{ Q,Q \} =0$.
Although the formulation in terms of a Poisson algebra is both
natural and transparent it does not help very much for
computing the cohomology. It only helps to see that the
cohomology $H^\bullet _D({\cal P})$ inherits a Poisson structure.

\vspace{5pt}
{\bf Remark}

2.1)
It is important here to remark that the construction of the
BRST complex is much more powerful in the general case
(\ref{struc}). There will always exist a $Q \in {\cal P}$
that squares to zero and that is uniquely defined up to
supercanonical transformations. This was proven in
[BV], [FF] and [HT].
The last paper also shows that
the BRST cohomology for regular first class constraints
equals the vertical cohomology of the constraint manifold.
The double complex structure is lost in general. This
is due to the presence of higher order structure
functions ([He]).

\vspace{5pt}

Non-regular constraints can be treated similarly,
enlarging ${\cal P}$. Again there exists a well defined
BRST charge and again its associated cohomology gives
the vertical cohomology of the constraint manifold
([FHST],[St]).
We will describe this in the next section
for (transitive) group symmetries, for which the BRST
complex is a double complex again.

\section{BRST complex for reducible symmetries}

In this section we also start with a Hamiltonian $S$-action
on a symplectic manifold $N$.
We will show how to obtain the de Rham cohomology from the
BRST complex for transitive $S$-actions.

We will assume that 0 is now only
a weakly regular value of the momentum map,
i.e., $Z$ is a submanifold but its
codimension is smaller than the number of constraints.
For this case, there
is a similar theorem on the existence of a BRST charge ([BV], [FHST]).
In this case there are relations among the constraints and
the differential $\delta$ has non-zero cohomology in
more than one degree now.
The constraints are called reducible.
To restore the resolution property (cohomology lives in one degree
only)
one enlarges the algebra
${\cal P}$ (introducing \lq anti-ghosts for anti-ghosts')
and modifies $\delta$
(resulting in what is called a Koszul-Tate resolution).

In the sequel we shall neglect $\delta$ and work
directly with the constraint manifold $Z$. Only considering
non-negative ghost degree is
convenient in the group case.
There is a double complex structure and $\delta$
is always constructed in such a way that the associated
spectral sequence degenerates.
We will denote the non-negative
degree part of ${\cal P}$ by ${\cal B}$.

 From now on,
we will always assume that the symplectic manifold $N$ is a
cotangent bundle $T^*M$ and that the constraints come
from  a Hamiltonian $S$-action that is induced by an arbitrary
action on $M$. $M$ is called the configuration space.
 $Z$ is then a subbundle of the vector bundle
$N$ over $M$ and the reduced phase space
$X=Z/S$ equals $T^*(M/S)$ (provided the quotient makes sense).
In the sequel we shall use $M$ instead of $Z$, because the action on
$Z$ comes from the one on $M$. It is in this
sense that we study the BRST complex
associated to an arbitrary Lie group action on an arbitrary
manifold.

In the sequel, we will assume that $S$ acts transitively on
$M$ (this property is crucial for cohomological field theories).
Let $S=H$ be
a finite dimensional Lie group with Lie algebra ${\got h}$,
$H_0$ a closed subgroup of $H$ and
let $M=H/H_0$.
We shall construct the BRST differential algebra
for the symmetry group $H$ acting on $M$. Consider the following
exact sequence of vector bundles

\begin{equation}\label{seq1}
       0 \rightarrow K_{\got h} \rightarrow V_{\got h}
         \rightarrow TM         \rightarrow 0
\end{equation}
where $V_{\got h}=M \times {\got h}$, which is mapped onto $TM$
using the infinitesimal action ${\got h} \rightarrow \Gamma (TM)$
followed by the evaluation map $M \times \Gamma (TM) \rightarrow
TM$. $K_{\got h}$ is the kernel of this map. Note that the
fibers of $K_{\got h}$ are isomorphic to ${\got h}_0$, but that
this bundle need not be trivial.
Note also that the corresponding sequence of sections of these bundles
is a sequence of ${\got h}-$modules by giving
$\Gamma(V_{\got h})=C^\infty(M) \otimes {\got h}$ the module
structure of its two components and $\Gamma(TM)$ the obvious one.
$\Gamma(K_{\got h})$ is then an ${\got h}-$module because it is
the kernel of a module map.

To apply (a geometrical version of the positive degree part of)
the [FHST] construction we need to distinguish two cases.
In the trivial case,
 $K_{\got h} \simeq M \times {\got h}_0$, the BRST
algebra equals

\begin{equation}\label{theirB}
      {\cal B}=S({\got h}^*_0) \otimes \Lambda({\got h}^*)
                               \otimes C^\infty(M)
\end{equation}

Elements in $S^1({\got h}^*_0)$ are called ghosts of ghosts
and have degree two.
If $K_{\got h}$ is not trivial, we need to find some
vector space ${\got h}_0'$
and one more exact sequence

\begin{equation}
      0 \rightarrow K_{{\got h}_0'} \rightarrow V_{{\got h}_0'}
        \rightarrow K_{\got h}      \rightarrow 0
\end{equation}
where $V_{{\got h'_0}}=M \times {\got h'_0}$ is mapped to $K_{\got h}$
using a map ${\got h_0}' \rightarrow \Gamma(K_{\got h})$
followed by the evaluation map.
If $K_{{\got h}'_0} \simeq M \times {\got h_1}$ for some
vector space ${\got h}_1$, then, by definition,

\begin{equation}
     {\cal B}=\Lambda({\got h^*}_1) \otimes S({\got h^*}'_0)
              \otimes \Lambda({\got h^*}) \otimes C^\infty(M)
\end{equation}
Elements in $\Lambda^k({\got h^*}_1)$ are of degree $3k$ and
are called ghosts of ghosts of ghosts.

If $K_{\got h'_0}$ is not trivial we proceed in the same way
finding another exact sequence for $K_{{\got h'}_0}$, etc.
To circumvent this definition process, we will use a
slightly altered definition. Let $SK^*_{\got h}$ be the
vector bundle with fibers the symmetric algebras of the duals
of the fibers of $K_{\got h}$, $\Gamma(SK^*_{\got h})$ its space
of sections.
The BRST algebra we will use in the sequel is

\begin{equation}\label{myB}
       {\cal B}= \Gamma(SK^*_{\got h}) \otimes \Lambda({\got h^*})
\end{equation}

Obviously, if $K_{\got h} \simeq M \times {\got h}_0$ then
$\Gamma(SK^*_{\got h}) \simeq S({\got h^*}_0) \otimes C^\infty(M)$
and (\ref{myB}) is isomorphic to (\ref{theirB}).
Furthermore, if $K_{\got h}$ has rank zero (free group action),
then ${\cal B}=H_\delta({\cal P})$, the BRST algebra of the previous
section.
Before we define the BRST operator on this algebra, let us introduce
a grading on ${\cal B}$. By definition, all elements of
       $ \Gamma(S^pK^*_{\got h}) \otimes \Lambda^q({\got h^*})$
have degree $2p+q$. This subspace will be denoted by
${\cal B}^{p,q}$.
We shall define the BRST operator $D$ now. Let
$\eta \in {\cal B}^{p,q}$. Then $D\eta= D_1 \eta + D_2 \eta$,
where $D_1 \eta \in {\cal B}^{p,q+1}$ and $D_2 \eta \in
{\cal B}^{p+1,q-1}$ are defined by

\[   D_{1}\eta ({\xi} _{1},...,{\xi}_{q+1})  =
{\displaystyle \mathop \sum_{i}}
(-1)^{i-1}{\xi} _{i} \cdot  \left [ \eta  \left ( {\xi} _{1},...,
             \widehat{{\xi}_{i}},...,{\xi}_{q+1}\right ) \right ]
+
\]
\begin{equation}\label{brstdif}
 +{\displaystyle \sum _{i<j}} \eta (\left [{\xi}_{i},{\xi}_{j}\right ],
                                       {\xi}_{1},..., \widehat{
  {\xi}_{i}},...,\widehat{{\xi}_{j}},..., {\xi}_{q+1}) (-1)^{i+j},
\end{equation}
\[  \! \!
   D_{2}\eta ({\xi}_{1},.., {\xi}_{q-1})
             ({\phi}_{1},..,{\phi}_{p+1})
    =
      {\displaystyle \sum_{i}}\eta\left (
           {\phi}_{i}, {\xi}_{1},..,{\xi}_{q-1}\right)
            \left( {\phi}_{1},..,\widehat{{\phi}_{i}},..,
                   {\phi}_{p+1}\right ),
\]
where all ${\xi} _{1},...,{\xi}_{q+1}$ are in $\got h$ and all
${\phi}_{1},..., {\phi}_{p+1} $ are in
$\Gamma(K_{\got h}) \subset C^\infty(M) \otimes {\got h}$.
$D_{1}$ is nothing but the differential of the standard complex
for the computation of the Lie algebra cohomology
of $\got h$  with values
in the ${\got h}$-module $\Gamma (SK^*_{\got h})$, and $D_2$ uses
the injection $\Gamma(K_{\got h}) \rightarrow C^\infty
(M) \otimes {\got h}$.

We are now going to exhibit a double complex structure on ${\cal B}$.
On the algebra ${\cal B}$, we define two degrees whose
sum will give our initial degree. The first degree of an element of
${\cal B}^{p,q}$ is $p$ and its second degree is $p+q$.
The first (respectively second) degree is preserved by $ D_{1}$
(respectively $D_{2}$) and is increased by one by $D_{2}$
(respectively $D_{1}$).

\begin{prop}
The two gradings defined above and the differentials
$D_{1}$, $D_{2}$ define a double complex structure on ${\cal B}$
whose associated total complex is the BRST complex.
\end{prop}

The reader be warned that this is another double complex
than the one of section 2.2. The relation is that in the case
that the rank of $K_{\got h}$ is zero, $D_2$ equals zero and
$D_1$ is the operator d acting on $H_\delta({\cal P})$.

{\bf Proof}.
We have to prove the following relations
$D_{1}^{2}=0$, $D_{2}^{2}=0$, $D_{1}D_{2}=-D_{2}D_{1}$.
It is a standard computation to establish the first one.
Let us prove the second one. Let $\eta$ be  in ${\cal B}^{p,q}$.
For all ${\xi} _{1},...,{\xi}_{q-2} \in {\got h}$ and all
${\phi}_{1},..., {\phi}_{p+2} \in
\Gamma(K_{\got h}) \subset C^\infty(M) \otimes {\got h}$, we have
\[         D_{2}^{2} \; \eta({\xi}_{1}, \ldots ,{\xi}_{q-2})
           ({\phi}_{1}, \ldots ,{\phi}_{p+2})=
\]
\[            = {\displaystyle
\sum _{i,j}}
     \eta \left ( {\phi}_{i}, {\phi}_{j}, \xi
_{1}, \ldots , {\xi}_{q-2}\right )\left ({\phi}_{1}, \ldots ,
\widehat{{\phi}_{i}}, \ldots ,
\widehat{{\phi}_{j}}, \ldots ,{\phi}_{p+2} \right )=0
\]
using the fact that this expression is symmetric in the ${\phi}_{i}$.

Let us now check the third relation. We compute separately
$D_{1}D_{2}$ and $D_{2}D_{1}$. They are both in ${\cal B}^{p+1,q}$.
Keeping the same notations, on one hand we have

\vspace{5pt}

$\begin{array}{l}
D_{1}D_{2}\eta ({\xi}_{1},...,{\xi}_{p})
({\phi}_{1},...,{\phi}_{q+1})  =            \\
={\displaystyle \sum_{i}} (-1)^{i-1} \left ({\xi}_{i} \cdot
\left ( D_{2}\eta
\left ( {\xi}_{1},..., \widehat {{\xi}_{i}},...,1{\xi}_{p}\right )
\right ) \right )
\left ( {\phi}_{1},...,{\phi}_{q+1}\right ) \\
 +{\displaystyle \sum_{i<j}}D_{2}\eta \left
( \left [ {\xi}_{i}, {\xi}_{j}\right ],..., \widehat{{\xi}_{i}},...,
\widehat {{\xi}_{j}},...,{\xi}_{p} \right )
\left (  {\phi}_{1},...,{\phi}_{q}\right ) (-1)^{i+j} \\
= {\displaystyle \sum_{i,k}}{\xi}_{i} \left
( \eta \left ( {\phi}_{k},{\xi}_{i},... \widehat{{\xi}_{i}}...
{\xi}_{p} \right )\left ( {\phi}_{1},..., \widehat{{\phi}_{k}},...,
{\phi}_{q+1}\right ) \right ) (-1)^{i-1}\\
- {\displaystyle \sum_{i,k}} D_{2}\eta \left ( {\xi}_{1},...,
\widehat{{\xi}_{i}},...,{\xi}_{p}\right ) \left ( {\phi}_{1},...,
\left [ {\xi}_{i},{\phi}_{k}\right ],..., {\phi}_{q+1} \right )
(-1)^{i-1}\\
+ {\displaystyle \sum_{i<j,k}}\eta
\left ( {\phi}_{k}, \left [ {\xi}_{i},
{\xi}_{j}\right ], {\xi}_{1},..., \widehat{{\xi}_{i}},...,
\widehat{{\xi}_{j}},...,{\xi}_{p} \right ) \left ( {\phi}_{1}, ...,
\widehat {{\phi}_{k}},...,{\phi}_{q+1}\right )(-1)^{i+j}\\
= {\displaystyle \sum_{i,k}}{\xi}_{i} \left
( \eta \left ( {\phi}_{k},{\xi}_{i},..., \widehat{{\xi}_{i}},... ,
{\xi}_{p} \right )\left ( {\phi}_{1},...,
\widehat{{\phi}_{k}},..., {\phi}_{q+1}\right ) \right ) (-1)^{i-1}\\
- {\displaystyle \sum_{i,k}}
\eta \left (\left [ {\xi}_{i}, {\phi}_{k} \right ], {\xi}_{1},...,
\widehat{{\xi}_{i}},...,{\xi}_{p}\right ) \left ( {\phi}_{1},...,
\widehat{{\phi}_{k}}..., {\phi}_{q+1} \right ) (-1)^{i-1}\\
- {\displaystyle \sum_{i,j \neq k}}
\eta \left ({\phi}_{j}, {\xi}_{1},...,
\widehat{{\xi}_{i}},...,{\xi}_{p}\right ) \left ( {\phi}_{1},...,
\widehat{{\phi}_{j}}, ...\left [ {\xi}_{i}, {\phi}_{k} \right ],
..., {\phi}_{q+1} \right ) (-1)^{i-1}\\
+ {\displaystyle \sum_{i<j}}\eta \left ( {\phi}_{k}, \left [ {\xi}_{i},
{\xi}_{j}\right ], {\xi}_{1},..., \widehat{{\xi}_{i}},...,
\widehat{{\xi}_{j}},...,{\xi}_{p} \right ) \left  ( {\phi}_{1}, ...,
\widehat{{\phi}_{k}},..., {\phi}_{q+1}\right )(-1)^{i+j}\\
\end{array}$

\vspace{5pt}

On the other hand,

\vspace{5pt}

$\begin{array}{l}
D_{2}D_{1}\eta({\xi}_{1},...,{\xi}_{p})({\phi}_{1},...,{\phi}_{q})
 = \\
= \sum_k
D_{1}\eta({\phi}_{k},{\xi}_{1},...,{\xi}_{p})
({\phi}_{1},..., \hat{{\phi}_{k}},...,{\phi}_{q}) \\
= {\displaystyle \sum_{i,k}} \left ( {\xi} _{i} \cdot \left ( \eta
\left ({\phi}_{k}, {\xi}_{1},..., \widehat {{\xi}_{i}},...{\xi}_{p}\right
)\right ) \right )
\left ( {\phi}_{1},...,\widehat {{\phi}_{k}},...{\phi}_{q+1}\right )
(-1)^{i}\\
 + {\displaystyle \sum_{k}}  {\phi} _{k} \cdot \left ( \eta
\left ({\xi}_{1},...{\xi}_{p}\right )\right )
\left ( {\phi}_{1},...,\widehat {{\phi}_{k}},...{\phi}_{q+1}\right ) \\
 + {\displaystyle \sum_{k,i<j}}
\eta \left (\left [ {\xi}_{i},{\xi}_{j}\right ], {\phi}_{k}, {\xi}_{1},...,
\widehat{{\xi}_{i}},...,
\widehat{{\xi}_{j}},...,{\xi}_{p} \right ) \left ( {\phi}_{1}, ...,
\widehat{{\phi}_{k}},..., {\phi}_{q+1}\right )(-1)^{i+j}\\
+ {\displaystyle \sum_{k,i}}
\eta \left (\left [ {\phi}_{k},{\xi}_{i}\right ],
{\xi}_{1},..., \widehat{{\xi}_{i}},...,{\xi}_{p} \right ) \left ( {\phi}_{1},
...,
\widehat{{\phi}_{k}},..., {\phi}_{q+1}\right )(-1)^{i}\\
\end{array}
$

\vspace{5pt}

As the second term equals zero,  and the first term can be expanded
further into two terms, it is clear that $D_{1}D_{2}=
- D_{2}D_{1}$.  $\Box$

\vspace{10pt}

According to [FHST], the cohomology of $({\cal B},D)$ is
isomorphic to the de Rham cohomology of $M$,
$H_D({\cal B}) \simeq H(M)$.
However, we need to prove this, since our complex is slightly
different from the one in [FHST].
\begin{thm}
        Let ($\Omega(M),{\rm d}$) be the de Rham complex associated
	to $M$. The map
  \begin{eqnarray}\label{fimap}
  \Phi : \Omega^k(M)=\Gamma(\Lambda^kT^*M) &\rightarrow&
           \Gamma(S^0K^*_{\got h}) \otimes \Lambda^k
                                           {\got h^*}
          \subset {\cal B}  \\ \nonumber
  \Phi (\eta)(\xi_1, \ldots ,\xi_k) &=& \eta(V_{\xi_1}, \ldots
             ,V_{\xi_k})
  \end{eqnarray}
induces an isomorphism of cohomologies.
More precisely, the spectral sequence associated to the
double complex (${\cal B},D$) degenerates at $E_2$ and the
de Rham complex is isomorphic with the $E_1-$term $H_{D_2}({\cal B})$.
\end{thm}

{\bf Proof}.
A simple computation shows that $D_1 \circ \Phi = \Phi \circ {\rm d}$
and that $D_2 \circ \Phi =0$. We will show that
Im($\Phi$) equals the cohomology of $D_2$. Using theorem \ref{dc2}
and the injectivity of (\ref{fimap}) will then finish the proof.

Choose an inner product on ${\got h}$. If $\iota$ denotes the
map $\Gamma(K_{\got h}) \rightarrow C^\infty(M) \otimes
{\got h}$, then this inner product, which induces a metric on
$M \times {\got h}$, gives a map $\pi:C^\infty(M) \otimes {\got h}
\rightarrow \Gamma(K_{\got h})$, such that $\pi \circ \iota$ is the
identity and $\iota \circ \pi$ is a projection.
Let us denote the ring $C^\infty(M)$ by $A$ and
the dual of the projection $\iota \circ \pi$ by $\rho:
A \otimes {\got h^*} \rightarrow A \otimes {\got h^*}$. The
$A$-module $A \otimes {\got h^*}$ splits in a direct sum
of two $A$-modules, the image $P_1$ of $\rho$ and its complement, the
image $P_2$ of $1-\rho$.
So $\rho$ is the projection operator on $P_1 \oplus P_2$, equal to
{\bf 1} (the identity) on the first term and equal to zero on the
second term. In the sequel, $\rho$ will also denote the extension
of $\rho$ to a derivation on
\[ A \otimes \Lambda{\got h^*} = \Lambda_A(P_1 \oplus P_2) \simeq
    \Lambda_AP_1 \otimes_A \Lambda_AP_2.
\]
Thus $\rho$ can be written as ${\bf 1} \otimes 1$, the identity
on $P_1$, extended as a derivation.
${\cal B}$ can be written
as $\Gamma(SK^*_{\got h}) \otimes_A \Lambda_AP_1 \otimes_A
\Lambda_AP_2$.
Note that $P_1$ is isomorphic to $\Gamma(K_{\got h})$, that
Im($\Phi$) is isomorphic to $1\otimes 1 \otimes \Lambda_AP_2$ and
that $D_2$ vanishes on this subalgebra. We will show now that
$D_2$ has trivial cohomology on the remaining factor
$\Gamma(SK^*_{\got h}) \otimes_A \Lambda_AP_1$.
Recall that
$D_2$ was defined by ($\eta \in \Gamma(S^pK_{\got h}) \otimes
\Lambda^q{\got h^*})$
\[  \! \! \! \! (D_2\eta)(\xi_1, .., \xi_{q-1})(\phi_1, .., \phi_{p+1})
  =\sum_j \eta(\iota(\phi_j), \xi_1, .., \xi_{q-1})
              (\phi_1, ..,\hat{\phi_j},..,\phi_{p+1})
\]
Let us define $C:{\cal B}^{p,q} \rightarrow {\cal B}^{p-1,q+1}$
by (for $p>0$)
\[
   (C\eta)(\xi_1,..,\xi_{q+1})(\phi_1,..,\phi_{p-1})=
\]
\[ =\sum_i
     (-1)^{i+1} \eta(\xi_1,..,\hat{\xi_i},..,\xi_{q+1})
                (\pi(\xi_i),\phi_1,..,\phi_{p-1})
\]
A straightforward computation shows that
\begin{equation}\label{chom}
             D_2 \circ C +
         C \circ D_2 = {\bf 1} \otimes 1 + 1 \otimes \rho,
\end{equation}
where {\bf 1} denotes the derivation coming from the
identity on ${\cal B}^{1,0}$.
Note that the second part is in fact the identity (extended as a
derivation) on $\Lambda_AP_1$.
So, from (\ref{chom}) we conclude
that $D_2$ has only non-zero cohomology in degree $(0,q)$ (the
only degrees for which $C$ is not defined).
Finally, if we define $C$ to be zero for degree $(0,q)$ elements,
then (\ref{chom}) remains true (the first term on the rhs is of
course zero in degree $p=0$) and it is easy to derive
that in this degree ker($D_2$)=Im($\Phi$), which proves
\[
  H_D({\cal B}) \simeq H_{D_1}(H_{D_2}({\cal B})) =
    H_{D_1}({\rm Im}(\Phi))
    \simeq H(M). \; \; \; \Box
\]

\chapter{BRST model for Cohomological Field Theories}

In this chapter we present a finite dimensional model for
Cohomological Field Theories.
This model will be derived from constructions in the
previous chapter by choosing an appropriate symmetry group.
We will describe how
path integrals look like in this model and identify the
partition function with the Mathai-Quillen representative for the
equivariant Thom class.
This analysis of path integrals uses the Fourier transform of
differential forms as described in
chapter one and was published in [Ka1].

\section{Introduction to the model}

Because in field theory one uses actions rather than Hamiltonians,
we will have to work with the path space of the configuration space
rather than with the configuration space itself. So from now on,
$M$ will represent a space of paths of some field theory.
The construction of the BRST complex is the same in both
cases, so that we can use the results of chapter two.

In Cohomological Field Theories, field theories with actions
${\cal S}=0$, the symmetry group is Diff$(M)$, the set of all possible
transformation of the fields (a field is just an element
of $M$). The stabilizer at each point is isomorphic to the
subgroup of diffeomorphisms leaving one point fixed.
Thus we are really in a $M=H/H_0$ situation of the previous
chapter, though there we took $H$ to be finite dimensional
to avoid analytical complexities.

In the most interesting Cohomological Field Theories there
is an additional symmetry group $G$ (gauge symmetries).
Of course, Diff$(M)$ contains already all possible
diffeomorphisms, so at first it is not obvious
what is gained by introducing $G$.
However, for field theories with actions that are only
$G$-invariant, the dynamics is on $M/G$ rather than on $M$ (like
the YM example in the introduction of this thesis).
Also, the space of all fields is often contractible,
so from a mathematical point of view $M/G$ is a lot more
attractive to study.
We will achieve this by constructing the BRST complex
associated to a $G \ltimes {\rm Diff}(M)$ symmetry on $M$
and then restrict the algebra appropriately.
We will prove that this restricted complex can be mapped to
the Cartan model for equivariant cohomology of $(M,G)$,
inducing an isomorphism  of the  cohomologies.

As in chapter two, we shall use a finite dimensional
Lie group $H$ instead of Diff$(M)$
(for the case that $H$ is infinite dimensional we refer to [CK]).
$H$ is supposed to act
transitively on $M$, so that $M \simeq H/H_0$ for a certain
$H_0<H$. Furthermore, the symmetry group will be the semi-direct
product
$S=G \semi H$, for the direct product $G \times H$
does not provide an action on $M$ (at least, not of the
form $(g,h)\cdot x = g \cdot (h \cdot x)$). Let us assume, for
simplicity, that $G<H$,
then the product in $S$ is defined as
\begin{equation}
    (g,h) \cdot (g',h') = (gg',(g')^{-1}hg'h')
\end{equation}
It is easy to check that this product gives rise to a well defined
group action of $S$ on $M$.
The Lie bracket, associated to this product on $S$, reads
\begin{equation}
   [(\xi,\nu),(\xi',\nu')]=([\xi,\xi'],[\nu,\xi']+
              [\xi,\nu']+[\nu,\nu'])
\end{equation}
Note that the Lie subalgebras $\{(0,\nu) \mid \nu \in {\got h} \}$ and
$\{(\xi,-\xi) \mid \xi \in {\got g} \}$
are invariant under the adjoint action,
whereas $\{ (\xi,0) \mid \xi \in {\got g} \}$ is not.
In fact, the first two subalgebras are commuting ideals whose
direct sum equals the whole algebra.
The reason for this is the existence of the following isomorphism
$G \ltimes H \rightarrow G \times H$,
$(g,h) \mapsto (g,gh)$.

\section{The model}

The aim of this section is to show that applying
chapter two to the symmetry group $S = G \semi H$, we obtain the
differential algebra $(B, \delta)$, where $B=W({\got g})
\otimes \Omega(M)$ and $\delta={\rm d}_W \otimes 1 + 1 \otimes
{\rm d}_M + \omega^a \otimes {\cal L}_a - \phi^b \otimes \iota_b$
(see section 1.2.3).
Restriction to a subalgebra will give the Cartan model for
equivariant cohomology.
To arrive at the differential $\delta$, we will need the splitting
(\ref{d1d2}) of d$_W \otimes 1$ into d$_1 + {\rm d}_2$.

The only difference with the previous chapter is that
we have a larger symmetry group $S=G \semi H$ here.
So, instead of the exact sequence (\ref{seq1}), we have
\begin{equation}
  0 \rightarrow K_{\got h} \oplus V_{\got g} \rightarrow
  V_{{\got g} \semi {\got h}} \rightarrow TM \rightarrow 0
\end{equation}
where $V_{\got g}$ denotes the trivial vector bundle $M
\times {\got g}$,
where the third arrow is given by $(x,(\xi,\nu)) \mapsto
V_{\xi +\nu}(x)$ and the second one consists of the imbedding
$K_{\got h} \rightarrow V_{\got h}$ and the map $V_{\got g}
\rightarrow V_{{\got g} \semi {\got h}}, \;\xi \mapsto (\xi,-\xi)$.
Thus the BRST algebra, associated to the $S$-action on $M$,
equals ${\cal B} =S({\got g^*}) \otimes \Gamma(SK^*_{\got h})
\otimes \Lambda({\got g^*}) \otimes \Lambda({\got h^*})$ and
is equipped with the following gradings
\begin{equation}
    {\cal B}^{p,q,r,s} = \left (S^p({\got g^*}) \otimes
                         \Lambda^q({\got g^*}) \right )
     \otimes \left ( \Gamma(S^rK^*_{\got h}) \otimes
                         \Lambda^s({\got h^*}) \right ).
\end{equation}

As in the previous chapter, we will use the map (\ref{fimap})
again, with ${\got g} \semi {\got h}$ instead of ${\got h}$.
If we followed the strategy of section 2.3, then
we would prove that the differential $D_2$ has non-zero
cohomology only in dimension $p+r=0$ (hence $p=r=0$) and end up with
the de Rham complex ($\Omega(M),{\rm d}$). However, since
we would like to end up with $W({\got g}) \otimes \Omega(M)$, we
will follow a slightly different strategy. We will split $D_2$
into two parts, $D_2^{\got g}$ and $D_2^{\got h}$, and only use
the latter to get rid of the $r$-grading and to keep the
$p$-grading. Furthermore, we will use that, applying (\ref{fimap}),
contraction and action of ${\got g}$ on $C^\infty(M) \otimes
\Lambda {\got h^*}$ corresponds to contraction and
Lie derivation of ${\got g}$ on $\Omega(M)$. This is immediate
after a glance at the definition (\ref{fimap}).

We shall compute the BRST differential now and identify
it with (\ref{e11}), using a map like (\ref{fimap}). Suppose $\eta \in
{\cal B}^{p,q,r,s}$, then, as in chapter two, $D\eta=D_1\eta +
D_2\eta$, where $D_1\eta \in {\cal B}^{p,q+1,r,s} \oplus
{\cal B}^{p,q,r,s+1}$ and $D_2\eta \in ({\cal B}^{p+1,q-1,r,s} \oplus
{\cal B}^{p+1,q,r,s-1}) \oplus {\cal B}^{p,q,r+1,s-1}$.
Thus we obtain operators
$D_2^{\got g}, D_2^{\got h}$, respectively given by projection
on the different homogeneous subspaces of ${\cal B}$.

\begin{thm}
$D_2^{\got h}$ commutes with $D_1+D_2^{\got g}$, hence $({\cal B},D)$
is a double complex with gradings $2p+q+r+s$ and $r$.
$D_2^{\got h}$ has cohomology only in degree $r=0$, so the
associated spectral sequence degenerates at $E_2$. Furthermore,
the map
\begin{equation}
   {\bf 1} \otimes \Phi :W({\got g}) \otimes \Omega(M)
         \rightarrow {\cal B},
\end{equation}
where $\Phi$ is the map (\ref{fimap})
and $W({\got g}) \otimes \Omega(M)$ is
equipped with differential (\ref{e11}),
${\rm d}_W \otimes 1 +1\otimes{\rm d} +
\omega^a\otimes {\cal L}_a -\phi^b\otimes \iota_b$, is a map of
differential algebras that maps $W({\got g}) \otimes \Omega(M)$
isomorphically onto the $E_1$ term $H_{D_2^{\got h}}({\cal B}).
$
\end{thm}

{\bf Proof}.
The first part of the theorem is straightforward. To prove that
$D_1 +D_2^{\got g}$ gives the differential (\ref{e11}), note that,
since $[{\got g},{\got h}] \subset {\got h}$, we can split
the Lie algebra cohomology operator $D_1$ into $D_1^{\got h}
+ D_1^{\got g}$, where $D_1^{\got h}$ is the
$\Gamma(SK^*_{\got h})$-valued Lie algebra cohomology
differential on ${\got h}$ (note that the action of
${\got h}$ on $S({\got g^*})$ equals zero because of the last
remark of section 3.1) and $D_1^{\got g}$ is the
$S({\got g^*}) \otimes \gamma(SK^*_{\got h}) \otimes \Lambda
{\got h^*}$-valued Lie algebra cohomology operator on ${\got g}$.
With these notations, the following observations are straightforward
and finish the proof of the theorem.

1) $D_1^{\got h}$ is probably the simplest. It is just the
Lie algebra cohomology operator on $\Lambda {\got h^*}$ with
values in $\Gamma(SK^*_{\got h})$, as defined in (\ref{brstdif}).
Under the map (\ref{fimap}) it
turns into the de Rham differential d on $\Omega(M)$.

\vspace{5pt}

2) $D_1^{\got g}$ is more complicated due to the fact that
${\got g}$ is embedded diagonally in ${\got g} \oplus {\got h}$.
This last fact implies that $D_1^{\got g}$ is not just the Lie algebra
cohomology operator on $\Lambda {\got g^*}$ with values in
$S({\got g^*})$, given by d$_2$ in (\ref{d1d2}),
but rather with values in
$S({\got g^*}) \otimes \Gamma (SK^*_{\got h}) \otimes
\Lambda{\got h^*}$. Thus, $D_1^{\got g}$ gives us not only
the d$_2$ part
of (\ref{d1d2}),
but also the $\omega^a \otimes {\cal L}_a$
part, after applying (\ref{fimap}).

\vspace{5pt}

3) $D_2^{\got g}$ consists of two parts, for the same reason.
A section of $V_{\got g}$ is mapped to itself, giving the d$_1$
part of (\ref{e11}) (see (\ref{d1d2})),
and to minus itself (viewed now as a section of
$V_{\got h}$), giving the $-\phi^b \otimes \iota_b$ part,
after applying (\ref{fimap}).

\vspace{5pt}

4) Finally, $D_2^{\got h}$ is used, as in section 2.3, to
identify the BRST complex (through a simple spectral sequence)
with the complex $W({\got g}) \otimes \Omega(M)$, inherited with
the differential (\ref{e11}). \hspace{3cm} $\Box$

\vspace{5pt}

The cohomology of the differential algebra above, the unrestricted
BRST algebra, is just $H(M)$
(see (\ref{isonr})). What we are really interested in is
the cohomology of the restricted algebra, $(S({\got g^*}) \otimes
\Omega(M))^G$. This is because we want to study quantum mechanics
on $M/G$ rather than on $M$ (which is an affine space, usually).
In the first chapter we saw that restriction gives the
Cartan model for equivariant cohomology.
So, starting with the Lie algebra cohomology complex, we
have finally
arrived at the Cartan model!

It turns out that from the point of view taken in this section,
the difference between the Weil model and the Cartan model is
just the choice of a basis in ${\got g} \oplus {\got h}$.
Indeed, if we define an action of $G \times H$ on $M$ by
$(g,h)\cdot x = h \cdot x$, then the previous construction
will give us the usual differential
${\rm d}_W \otimes1 + 1 \otimes {\rm d}$.
 From the isomorphism at the end of section 3.1 we see that both
group actions are in fact the same and that the difference in
the differentials comes from the choice of a basis to describe the
Lie algebra cohomology operator.

\section{Incorporation of anti-ghosts}

Up till now, we have constructed the following model for a
$G \semi {\rm Diff}(M)$-action on a space of fields $M$. The positive
degree part of the BRST algebra equals $(S({\got g^*}) \otimes
\Omega(M))^G$ and the BRST operator on this algebra equals
$s=1 \otimes {\rm d} - \phi^b \otimes \iota_b$.
In order to compute path integrals (representing physical
quantities) we have to choose
a (gauge fixing) function $F:M \rightarrow V$ (like the
functions $g^i$ at the beginning of chapter two), where $V$ is some
vector space of dimension $n$.
To respect the $G$-action, we require $V$ to
be a representation space for $G$ and $F$ to be equivariant.
In the case of a free $G$-action, $F$ would represent
a section of the associated vector bundle $M \times_G V$.

We will describe how to obtain a quantum action using this $F$
and we will prove that the final expression is the
Mathai-Quillen representative for the equivariant Thom class.
As in the introductory example of chapter two, let us introduce
some more variables:
\begin{equation}
    {\cal P}=\left(S({\got g^*}) \otimes \Omega(M) \otimes
             \Omega(V) \otimes \Omega(V^*) \right)^G
\end{equation}
We extend $s$ as follows. On $S({\got g^*}) \otimes
\Omega(M) \otimes \Omega(V)$ it is the differential of the
Cartan model for $(G,M \times V)$. On $S({\got g^*}) \otimes
\Omega(V^*)$ it is the Fourier transform (${\cal F}:\Omega(V)
\rightarrow \Omega(V^*)$) of $s$ restricted to $\Omega(V)$.
This extension coincides with the description in, e.g.,
[OSvB].

Let $z^i$ denote linear coordinates on $V$, $b_i$ the
dual ones on $V^*$. Furthermore, let $\psi^i$=d$z^i$ and
$\bar{\psi}_i$=d$b_i$, then $\iota_\xi(\psi^j)=\xi \cdot z^j$
(for all $\xi \in {\got g}$) and

\[{\cal F}(\iota_\xi (f \otimes \eta)) =
 \int f \; e^{i \;<b \mid z>} \otimes (\iota_\xi \; \eta)
e^{i\bar{\psi}_j \psi^j} dzd\psi \;  =
\]
\[ =-\int f \; e^{i\; <b \mid z>} \otimes (-1)^\eta \eta
(\iota_\xi e^{i\bar{\psi}_j \psi^j}) dzd\psi \;  =
\]
\[
=\int i\xi(z^j) f \; e^{i\; <b \mid z>} \otimes  \bar{\psi}_j\eta \;
e^{i\bar{\psi}_j \psi^j} dzd\psi \;  =
\]
\[=\xi(\sfrac{\partial}{\partial b_j}) \otimes \bar{\psi}_j
{\cal F}(f \otimes \eta)
= (-\sfrac{\partial}{\partial b_j} \otimes \xi \cdot \bar{\psi}_j)
{\cal F}(f \otimes \eta).
\]
for any $f \otimes \eta \in \Omega(V)$. Therefore, $s$ acts as
follows on $\Omega(V) \otimes \Omega(V^*)$

\begin{eqnarray}
s(z^i) = \psi^i  , & & s(\psi^i) = -\phi^a \otimes X_a \cdot z^i \nonumber \\
s(\bar{\psi}_i) = -b_i, & &
s(b_i) =
\phi^a \otimes X_a \cdot \bar{\psi}_i
\end{eqnarray}

Let us state all this in another way. If $D_M={\rm d}-\phi^b \iota_b$ is
the Cartan differential for $(G,M)$, $D_V$ is the one for $(G,V)$, then
\begin{equation}
      s=D_M + D_V + {\cal F} \circ D_V \circ {\cal F}^{-1}
\end{equation}
Using this one easily sees that integrating $\eta e^{i \; s(z^j
\bar{\psi}_j)}$ over $V^*$, where $\eta \in {\cal P}$ is any
$s$-closed element, one obtains $\eta' \in
\left(S({\got g^*}) \otimes \Omega(M) \otimes \Omega(V) \right)^G$
that is closed for $D_M +D_V$.
This is because the exponential indicates that we are really
Fourier transforming $\eta$ (see remark 1.5).
If we pull this Fourier transformed element back using
the map $F:M \rightarrow V$, we obtain an equivariantly
closed differential form on $M$.
In this way path integral quantization
combined with BRST theory gives rise to representatives
for equivariant cohomology classes.

\section{Path integrals and correlation functions}

The following proposition is an equivariant version of
(\ref{tc1}).
\begin{thm}\label{tc2}
The element
\begin{equation}\label{reptc2}
                   \sfrac{1}{(2\pi)^n}
\int_{V^*} e^{i \; s(<z-ib \mid \bar{\psi}>)} db \; d\bar{\psi}
= \sfrac{1}{(\sqrt{\pi})^n} e^{-z^2} \;
             \int e^{i \; <\psi \mid \bar{\psi}> +
                      i \; <\phi \cdot \bar{\psi} \mid \bar{\psi} >}
               d\bar{\psi}
\end{equation}
is a closed form in $(\Omega(V) \otimes S({\got g}^*))^G$ that
represents
the equivariant Thom class.
\end{thm}

${\bf Proof.}$
Essentially, the proof consists of showing that (\ref{reptc2}) is closed
under d$-\phi^a \otimes i_a$ and that its top form part
equals (\ref{reptc1}), as in the first example of section 1.4.2.
Note that the integrand above
can be written as a product $e^{i\; s <z \mid \bar{\psi}> \;
} \; e^{s<b \mid \bar{\psi}>}$, which means that we are Fourier
transforming $e^{s<b \mid \bar{\psi}>}$.
Because this form is obviously $s$-closed, the integral
itself is $D_V$-closed. Furthermore, its
top form part equals expression (\ref{reptc1}).
This can be verified by replacing $\phi$ by zero.
$\Box$
\vspace{5pt}

{\bf Remark}

3.1)
This representative is the same as the one constructed in [MQ].
Also in [AJ] the connection between this representative and
quantum actions of cohomological field theories is explained.

\vspace{5pt}

To end this section, we shall make some comments on the context
in which this theorem may be used.
Suppose that the $G$-action is free. We can pull back (\ref{reptc2})
by $F:M \rightarrow V$ to obtain a representative
(in the Cartan model) for the
Poincar\'e dual of $F^{-1}(0)/G$ in $M/G$.
Physicists use the expression
for the Poincar\'e dual to compute quantum correlation functions
of BRST invariant observables. We shall explain what
this means.

Denote $F^{-1}(0)/G \subset M/G$ by $X$ and let $i:X
\rightarrow M/G$ be the inclusion. Using the formula above, we
obtain a multilinear function
on $H^*(M/G)$ by integrating products of closed
differential forms over $M/G$. This polynomial, evaluated at a given
product of forms, is called a quantum correlation function.
BRST invariant observables for a topological quantum (field) theory
are just cohomology classes on the configuration space (which
equals $M/G$ in this case). So, mathematically, it is clear what
the correlation functions are. They are products of classes in
$H^*(M/G)$ (of total degree $\dim(X)$) integrated over $M/G$ using
the Poincar\'e dual of $X$.

Two problems remain. The first is that we would like to integrate
over $M$ rather than over $M/G$ (to incorporate non-free group
actions). Secondly, we still have to
apply the Weil homomorphism to get a differential form representative
instead of a Cartan model representative. BRST theory solves the
two problems together by introducing a dual Weil algebra, $W({\got g}^*)
=S({\got g}) \otimes \Lambda({\got g})$, generated by
$\bar{\phi}_a$ (degree -2) and $\bar{\omega}_a$ (degree -1)
($a=1, \ldots , \dim(G)$). The BRST differential extends as follows
\begin{equation}
s \; \bar{\omega}_a = -f^c_{ab} \phi^b \otimes \bar{\phi}_c , \hspace{20pt}
s \; \bar{\phi}_a = \bar{\omega}_a
\end{equation}

Furthermore, a Riemannian metric on $M$ and a non degenerate invariant
bilinear form on ${\got g}$ is used to obtain a linear
map $\nu: {\got g}^* \rightarrow \Omega(M)$ from the infinitesimal
action ${\got g} \rightarrow \Gamma(TM)$. Because we have chosen a
basis of ${\got g}$ we can define the 1-forms $\nu^a$ to be the
components of the map $\nu$.

To solve the two problems mentioned above, just add the following
factor to the integrand
\begin{equation}\label{eqn47}
\int_{W({\got g}^*)} e^{i \; s(\bar{\phi}_a \nu^a)}
      d\bar{\phi} \; d\bar{\omega}
\end{equation}
and perform an extra integration over the $\phi^a$. In [AJ] it is shown
that this is equivalent with multiplying by the vertical top form
and substituting the \lq Riemannian' curvature for $\phi$, thereby
solving the two problems together.
\vspace{10pt}

{\bf Remarks}

3.2) We are well aware of the fact that this last part is not at all
self contained. This is because we do not like rewriting parts of
[AJ]. Note that we do not need to calculate the precise expression
for the connection 1-form, because a change of $\nu$ can be absorbed
by a transformation of the variables $\bar{\phi}$ and $\bar{\omega}$,
of which the Jacobians cancel each other (as usual).

3.3) The final expression we obtain this way has also meaning in the
case the action of $G$ is not a free action. We still have a
polynomial on cohomology classes in the Cartan model expressed as
an integral over $M$. It is only more difficult to identify the
cohomology of the Cartan model with the de Rham cohomology of the
quotient space, since the latter is not well defined.
However, this is a serious case to investigate, since it is very
common in physics to have non free group actions (e.g., in TYMT, if
there are reducible connections).
In the next chapter we will make a start with this,
applying the model to the case of a symplectic manifold $M$ on
which $G$ acts Hamiltonially
and where $F$ is the momentum map $\mu$.

3.4) Combining expressions (\ref{reptc2}) and
(\ref{eqn47}), we get the quantum
action for TYMT as derived in [W1].

\chapter{Applications to symplectic geometry}

In this last chapter we will see that the integrals of
chapter three lead to a fixed point formula for
equivariantly closed forms. Applying this to
Hamiltonian circle actions we will show how to obtain
information on the ring structure of the cohomology of the
symplectic quotient. As an illustration we compute
the cohomology rings of all possible symplectic quotients
of ${\bf CP}^n$ and $S^1$.
Apart from the localization formula for equivariant forms on
manifolds with boundary, all the results here are also in
[Ka2]. Similar results were obtained (a few months earlier)
by Wu ([Wu]), using [W3].
Very recently, a beautiful generalization appeared in [JK].

\section{A fixed point formula}

Let $M$ be a compact manifold with a boundary $\partial M$. Let $\mu$
be a non-negative function on $M$ such that $\partial M =
\mu^{-1}(0)$. Suppose $M$ is equipped with a circle action
such that $\partial M$ is invariant and the fixed points
do not lie on the boundary. This implies
that all stabilizer groups of points on the boundary are
finite and thus $X:=\mu^{-1}(0)/S^1$ is an orbifold ([Sa]).
We will assume that the main stabilizer
is $\{ 1 \}$ to avoid extra factors in the formulas.
We shall denote by $V$ the generating vector field of the
circle action ($S^1={\bf R}/{\bf Z}$).

\subsection{Isolated fixed points}

We shall first assume that all fixed point are isolated, so
there are only finitely many of them.
Let $\alpha=\sum \alpha_j \phi^j \in {\bf R}[\phi] \otimes
\Omega(M)^{S^1}$ be an equivariantly closed differential form
on $M$, i.e., ${\rm d} \alpha_j = \iota_V \alpha_{j-1}$ for all
$j$. It is mapped to a closed differential form $r(\alpha)$ on
$X$ by
restriction to $\partial M$ followed by the map that gives the
equivalence between the Cartan model and differential forms
on the quotient. This will be explained in more
detail in the next section.
The integral of $r(\alpha)$ over $X$ can be expressed as a
summation over the fixed point set $F \subset M$
by means of the following formula.

\begin{thm}\label{fp}
Let $M$ be an $S^1$-manifold with an invariant boundary
$\partial M$ and isolated fixed points $F$. Let $\alpha$
be an equivariantly closed form on $M$ of total degree
$2n-2={\rm dim}(M)-2={\rm dim}(X)$. Then
\begin{equation}\label{4}
\int_{\partial M/S^1} r(\alpha) =  \sum_{P \in F}
                   \frac{\alpha^{(0)}_{n-1}(P)}
                        {\prod \; n_i(P)}
\end{equation}
where $\Pi n_i(P)$ is the product of the weights of the circle action
linearized around the fixed point $P$
and the superscript $(0)$ denotes the fact that the
differential form $\alpha_{n-1}$ is of degree zero.
\end{thm}

{\bf Proof}. Let $M_\epsilon$ denote the manifold $M$ where
small neighbourhoods around the fixed points are removed.
On $M_\epsilon$ the circle action is locally free so we
can find a connection 1-form $\theta$ on $M_\epsilon$.
The theorem can be proved by using
the following integral (of chapter three)
\begin{equation}\label{complint}
      \frac{1}{\sqrt{\pi}} \int_{M_\epsilon}\int{\rm d}\phi
      (\sum \alpha_j \phi^j)(e^{-\mu^2} {\rm d}\mu)
     \int_{W({\got g^*})} e^{i\bar{\omega}\theta}
          e^{i\bar{\phi}({\rm d}\theta - \phi)}.
\end{equation}
Performing the integration over $\phi$, $\bar{\phi}$ and
$\bar{\omega}$, we obtain
\begin{equation}
            \frac{1}{\sqrt \pi}
    \int_{M_\epsilon} e^{-t^2\mu^2} {\rm d} (t\mu) \wedge \theta
                  \wedge \sum \alpha_j ({\rm d} \theta)^j
\end{equation}
where we substituted $t\mu$ for $\mu$ ($t \in {\bf R}$).
Computing the limit $t \rightarrow \infty$ in two different
ways one obtains the identity (\ref{4}). This proof was given
in [Ka2]. A more direct proof goes as follows.

Since ${\rm d}(\theta \wedge \sum \alpha_j ({\rm d}
\theta)^j)= \iota_V (\theta \wedge \sum \alpha_j ({\rm d} \theta )
^{j+1})$, the top form part on both sides must be zero.
Therefore,
\begin{equation}\label{27}
        0=
                   \int_{M_\epsilon} {\rm d}
              (   \theta \wedge \sum  \alpha_j ({\rm d}\theta)^j)
       =            \int_{\partial M_\epsilon}
              (   \theta \wedge \sum  \alpha_j ({\rm d}\theta)^j)
\end{equation}
One part of the boundary of $M_\epsilon$ is the boundary of $M$,
hence part of the rhs equals
\begin{equation}\label{16}
\int_{\mu^{-1}(0)} \theta \wedge \sum \alpha_j
                                  ({\rm d}\theta)^j
\end{equation}
It is important to remark that
the integrand of (\ref{16}) equals $\theta \wedge r(\alpha)$
(this will be explained in the next section).
Integration over the orbits of the circle action gives us
the lhs of the theorem.

The other part of the boundary consists of boundaries of
small neighbourhoods around the isolated fixed points.
We will choose local coordinates around each fixed point
to compute (\ref{27}).
Identify an open neighbourhood of some fixed point $P$ with
${\bf R}^{2n}$ such that $S^1$ acts linearly with weights $n_i(P)$,
$i=1, \ldots, n$. Using linear coordinates $x^i, y^i$ we can
choose
\begin{equation}\label{conn}
\theta = \frac{\sum n_i (x^i {\rm d} y^i - y^i {\rm d} x^i)}
              {\sum n_i^2 ((x^i)^2 +(y^i)^2)} \cdot
              \frac{1}{2\pi}
\end{equation}
Using smooth partitions of unity this connection has an
extension to a smooth connection on $M_{\epsilon}$.
The only term of the integrand
in (\ref{27}) that gives a non-zero contribution
is the term with the highest singularity, $\theta \wedge
({\rm d} \theta)^{n-1}$. Integrating this term over some area,
surrounding the origin, gives $1/\prod \; n_i(P)$
(the most convenient way to do the calculation is to use
polar coordinates in each ${\bf R}^2$).
Noticing that the coefficient $\alpha_{n-1}$ is a zero form,
we finally obtain the rhs of the formula in the theorem.

\vspace{5pt}

{\bf Remark}

4.1)
Let us apply the theorem to a Hamiltonian circle action with
momentum map $\mu$. Let $M$ be the manifold given by
$\mu \geq 0$. The formula expresses integrals over the reduced
phase space $X=\mu^{-1}(0)/S^1$
as a sum over fixed points $P$
for which $\mu(P)>0$. Of course, we get a similar formula
for $\mu \leq 0$.
Combining these formulas and
using the equivariantly closed form
$\alpha= (\sigma - \mu \phi)^j \phi^{n-j-1}$, we get the following
series of identities
\begin{equation}
          \sum_{P \in F} \frac{\mu(P)^j}{\prod n_i(P)} =0
       \; \; \;    (j=0,1, \ldots ,n-1),
\end{equation}
where $F$ is now the set of all fixed points of the Hamiltonian
circle action.
These identities also follow directly from the Duistermaat-Heckman
formula ([DH]), see, e.g., [Wu], corollary 3.3.

\subsection{Non-isolated fixed points}

Going through the proof once more, one sees that the theorem
can be extended to the case of non-isolated fixed points.
Suppose $F$ is a union of connected submanifolds $F_i^{(2k)}$, where
the upper index denotes the dimension of the component.
Let $X$ again denote the quotient $\partial M/S^1$.
If the normal bundle of $F$ in $M$ is trivial, we can use
local analysis as above and the
formula of the theorem generalizes to
\begin{equation}\label{genfpf}
\int_X r(\alpha) = \sum_{F^{(2k)}_i \subset F}
                   \frac{
                   \int_{F_i^{(2k)}} \alpha_{n-k-1}^{(2k)}}
                   {\prod \; n_j(F_i^{(2k)})}
\end{equation}

However, if the normal bundle of $F$ is non-trivial, the rhs of
(\ref{genfpf}) is only the first term of a much more complicated
formula involving $\alpha_j$ for $j>n-k-1$ and characteristic
classes of the normal bundle. We shall derive this formula here,
using localization for integrals of equivariantly closed
forms ([BGV]).

Notice that we are after an expression for $\int \theta
\wedge \sum \alpha_j ({\rm d} \theta)^j$ around $F_i^{(2k)}$
as an integral over $F_i^{(2k)}$. Proofs of equivariant
localization provide such an expression. In fact, the form
$\theta \wedge \sum \alpha_j ({\rm d} \theta)^j$ appears
explicitly in [DH] (Addendum, (2.6)-(2.11)).
Replacing their factor $(-1)^ke^{J_X}\frac{\sigma^{n-k}}{(n-k)!}$
by $\alpha_{k-1}$, their
local computation gives us
(together with the localization formula in [BGV])

\begin{equation}\label{nifpf}
  \int_X r(\alpha) = \sum_{F_i^{(2k)} \subset F}
                     \int_{F_i^{(2k)}}
      \frac{\sum \alpha_j \phi^{j+1}}{\epsilon_i(\phi)}
\end{equation}
where $\epsilon_i(\phi)$ is the equivariant Euler class of
the normal bundle of $F_i^{(2k)}$ in $M$. Note that the rhs
should be independent of $\phi$ and that substituting
$\epsilon_i(\phi)= \phi^{n-k} \prod_{j=1}^{n-k} n_j(F_i^{(2k)})
+ l.o.t.$ gives the rhs of (\ref{genfpf}) as a starting term.

\vspace{15pt}

It's appropriate to say something here on Witten's
non-abelian localization ([W3]), since his methods also
produce a (different) proof of the theorem above.

Let $G$ act on a manifold $M$
of dimension $m$. Let ($(S({\got g}^*)
\otimes \Omega(M))^G$, $D$) be the Cartan model for the
equivariant cohomology of $M$. Define $S({\got g}^*)$-valued
integration over $M$ of
an equivariant form $\alpha=\sum_i p_i \otimes \alpha^{(i)}$,
where $\alpha^{(i)} \in \Omega^i(M)$ and $p_i \in S({\got g^*})$
as follows
\begin{equation}
\int_M \alpha = \int_M \sum_i p_i \otimes \alpha^{(i)}
              = p_m \cdot \int_M \alpha^{(m)}
\end{equation}
It is important to note  that this equivariant integration
gives zero on $D$-exact forms, so that it is a well defined
operation on equivariant cohomology classes.
Now, let $\lambda \in \Omega^1(M)^G$ and let $\alpha$ be a closed
equivariant form. Since ${\rm exp}(itD(\lambda))-1$ is $D$-exact,
the forms $\alpha$ and $\alpha \wedge {\rm exp}(itD(\lambda))$
are in the same class, so
\begin{equation}\label{12}
\int_M \alpha = \int_M \alpha \wedge e^{itD(\lambda)}
\end{equation}
This is the key of Witten's non-abelian localization.
Since $D(\lambda)$ has a component $\iota_V \lambda \in
\Omega^0(M)$, the rhs of (\ref{12}) localizes at the critical
points of $\iota_V \lambda$ for $t \rightarrow \infty$.
Of course, the integral does not depend on $t$, so the
answer can exactly be computed using stationary phase approximation.
Witten integrates both sides of (\ref{12}) over $\got g$,
interpreting the answers in terms of distributions.
In [Wu] the computation is carried out explicitely for
Hamiltonian circle actions
and some special choise of $\lambda$ and $\alpha$.
Formulas are obtained
similar to (\ref{4}).

\vspace{5pt}

{\bf Remark}

4.2) We can use (\ref{12}) to prove the following general
localization formula for manifolds with boundary (under the
same assumptions as above), extending the one of
[AB] and [BGV].
\begin{equation}\label{glf}
        \int_M \sum \alpha_i \phi^i =
      \sum_{F_i \subset F}  \int_{F_i} \frac{\sum \alpha_i \phi^i}
                                {\epsilon_i(\phi)}
                   -
       \int_{\partial M/S^1} \sum_{i,k} \tilde{\alpha}_i
         ({\rm d}\theta)^k \phi^{i-k-1}
\end{equation}
where ${\rm d}\theta$ represents the curvature class of
$\partial M \rightarrow X$ on the base space, and
$\tilde{\alpha}_i$ is the form on $X$ that comes from the
basic form $\alpha_i-\theta \wedge \iota_V \alpha_i$ on
$\partial M$. The proof is almost the same as in [BGV].
the only difference is an extra boundary term
$\int_{\partial M} \lambda \alpha \wedge \frac{e^{itD\lambda}-1}
{D\lambda}$ that appears in (\ref{12}).
Using this localization formula, we can derive the fixed point
formula (\ref{4}) immediately. Taking $\alpha$ homogeneous
of degree $2n-2$, the lhs vanishes and the rhs gives the
formula.

So, if people would have looked at localization formulas on
manifolds with boundaries, the fixed point formula for
integrals over symplectic quotients of circle actions
would have been found much earlier.

\section{Cohomology of symplectic quotients}

In this section we will explain how the fixed point formula
can be used to compute the ring structure on the cohomology
of symplectic quotients.
We will start with some results for the Cartan model of
equivariant cohomology.
For the following special case the equivariant cohomology $H^*_G(M)$
can be computed (as a vector space).

\begin{prop}\label{prop3.1}
Let $M$ be a manifold with a compact $G$-action for which the odd
Betti numbers are zero. Then the map
\begin{equation}\label{dcc2}
    H^l_G(M) \rightarrow \bigoplus_p (S^p({\got g^*})^G
                    \otimes H^{l-2p}(M)),
\end{equation}
which associates to an equivariant cohomology class its
part involving its highest form degree,
is well defined and surjective for every $l \geq 0$.
Moreover, {\rm dim}$H^l_G(M)=\sum_p {\rm dim} S^p({\got g^*})^G
\cdot {\rm dim} H^{l-2p}(M)$.
\end{prop}

{\bf Proof.}
The proof is based on the fact that the Cartan model actually
has a double complex structure. To see this, note that
$1 \otimes {\rm d}$ and $\phi^b \otimes \iota_b$ (super)
commute because of the $G$-invariance. Furthermore, note that
$\phi^b \otimes \iota_b$ increases the polynomial degree by
one and does not change the sum of the polynomial degree and the
form degree. For $1 \otimes {\rm d}$, vice versa.

Since the difference of the two degrees is the form degree and
since the odd Betti numbers vanish, we can apply theorem
\ref{spseq1}, which proves the proposition.

\vspace{5pt}

{\bf Remark}

4.3) What this proposition in fact says is that, for the special case
of vanishing odd Betti numbers, we know the dimensions of all
the $H^l_G(M)$ as well as the top form parts
(provided we know both $M$ and $G$ well enough). Reconstructing
from an invariant
top form part a D-closed element is called the zig-zag
construction ([BT]). Obviously it is non-unique.

\vspace{5pt}

 From chapter one we recall that if $M$ is a
symplectic manifold, then the $G$-action
in proposition \ref{prop3.1} is Hamiltonian.
In fact, the proposition is true in a more general setting.

\begin{thm}\label{thmKi}
{\rm ([Ki], [Gi])}.
Let $G$ be a connected compact Lie group acting Hamiltonially
on a compact symplectic manifold $M$. Then the
map
\begin{equation}\label{dcc3}
    H^l_G(M) \rightarrow \bigoplus_p (S^p({\got g^*})^G
                    \otimes H^{l-2p}(M))
\end{equation}
which associates to an equivariant cohomology class its part involving
the highest form degree is well
defined and surjective for every $l \geq 0$.
Moreover, {\rm dim}$H^l_G(M)=\sum_p {\rm dim} S^p({\got g^*})^G
\cdot {\rm dim}H^{l-2p}(M)$.
\end{thm}

The proof of this theorem is more complicated than the proof of
proposition \ref{prop3.1}. One can find it in [Ki] and [Gi].
The usefulness of this theorem is obvious. It says that
the spectral sequence associated to the Cartan model degenerates,
so that one can use the zig-zag construction to
construct equivariant representatives from invariant polynomials
on $\got g$ with values in
the space of closed invariant differential forms.

As an example we take the
constant polynomial equal to the symplectic 2-form $\sigma$.
It is closed for $1\otimes {\rm d}$ and from the theorem we
know that $\phi^b \otimes \iota_b \sigma$ must be $1 \otimes$d-exact.
Actually, this is precisely the condition that the action is
Hamiltonian. Thus we see that $\sigma$ corresponds to the
equivariant representative $\sigma - \phi^b \otimes \mu_b$, where
$\mu_b$ are the components of the momentum map.

Let us assume now that $G$ acts
locally free on $Z=\mu^{-1}(0)$ and denote the quotient by $X$.
Consider the following sequence of maps, starting with
equivariant forms on $M$ and ending in the
differential forms on the quotient $X$.
\begin{eqnarray}\label{psimaps}
(S({\got g^*}) \otimes \Omega(M))^G \stackrel{\Psi_1}{\rightarrow}
(W({\got g}) \otimes \Omega(M))_{\rm basic}
              \stackrel{\Psi_2}{\rightarrow}
              (W({\got g}) \otimes \Omega(Z))_{\rm basic}
\end{eqnarray}
\[
     \stackrel{\Psi_3}{\rightarrow} \Omega(Z)_{\rm basic}
               \rightarrow \Omega(X)
\]
where $\Psi_1$ is the map $\Psi$ of section 1.2.3, $\Psi_2$ is just
restriction to $Z$ and  $\Psi_3 = c_\theta
 \otimes 1$ is the Chern-Weil homomorphism (section 1.2.2).
We shall denote the composition
of these maps by (this notation was already introduced in the
previous section to state the fixed point formula)
\begin{equation}
r: (S({\got g^*}) \otimes \Omega(M))^G \rightarrow \Omega(X)
\end{equation}

Kirwan ([Ki]) has shown that the map $\Psi_2$ induces a
surjective map on the level of cohomology. Since the other
maps induce isomorphisms of cohomologies, the induced map
\begin{equation}\label{rstreep}
  \bar{r} : H_G(M) \rightarrow H(X)
\end{equation}
is an epimorphism. This is essential for the cohomology computations
later on.

We shall explicitely describe (\ref{psimaps}) for
circle actions. Let $V$ be the generating vector field and
let $\theta$ be a connection one form on $Z$,
that is, $\iota_V \theta =1$. Furthermore,
let $\sum \alpha_j \phi^j \in {\bf R}[\phi] \otimes \Omega(M)^{S^1}$
be some equivariant differential form on $M$. Then
\begin{equation}\label{rmap}
   r(\alpha) = \sum \alpha_j ({\rm d}\theta)^j -
               \theta \wedge \sum (\iota_V \alpha_j)({\rm d}\theta)^j
\end{equation}
which is a basic form on $Z$ and hence descends to
the symplectic quotient $X$.
Note that to obtain the first term on the rhs we just substituted
the curvature form d$\theta$ for $\phi$ in $\alpha$.

To conclude this section we explain how one computes the
ring structure on $H(X)$.
The map (\ref{rstreep})
is a ringhomomorphism and therefore the ring structure
on $H(X)$ comes from the ring structure on $H_G(M)$ by dividing
out the kernel ker($\bar{r}$).
In principle, this kernel can be computed using
Poincar\'e duality for orbifolds ([Sa]).
Since $X$ is compact,  integration over
$X$ can be regarded as a non-degenerate bilinear form on
the cohomology.
If we know how to integrate elements coming from $H_G(M)$
we can compute ker($\bar{r}$)
in the following way. A closed equivariant form $\alpha$
represents a class in
ker($\bar{r}$) iff
\begin{equation}\label{pd}
   \int_X r(\alpha) \wedge r(\beta) = 0
\end{equation}
for all closed equivariant forms $\beta$.
Using our fixed point formula we can compute the
lhs and thus determine which $\alpha$ are in the kernel of $r$.
We will do this computation now for a series of
examples.

\section{The {\bf CP}$^n$-case}

In this section we will give a series of examples
with $M={\bf CP}^n$ and $G=S^1$. We shall
compute the cohomology rings of all
symplectic quotients of ${\bf CP}^n$ and $S^1$.

\subsection{The circle action}

Let $(z_0:\ldots:z_n)$ denote homogeneous coordinates on
{\bf CP}$^n$. Every (linear) circle action on ${\bf CP}^n$
can be characterized by an $(n+1)-$tuple of integers $m_i$,
such that
\begin{equation}\label{lam}
\lambda \cdot (z_0:\ldots:z_n)=
(\lambda^{m_0}z_0:\ldots : \lambda^{m_n}z_n)
\end{equation}
Note that if the $m_i$ are shifted all by the same integer,
then the action does not change.
This action is Hamiltonian for the standard symplectic form
on {\bf CP}$^n$. The requirement of isolated fixed points is
equivalent with all $m_i$ being different. The fixed point set $F$ in
this case consists of
the points $P_i=(0: \dots:1:\ldots :0)$, where the 1 is on the
$i-$th place.
As Hamiltonian (momentum map) we take
\begin{equation}\label{ham}
           \mu_\nu : {\bf CP}^n \rightarrow {\bf R}, \;\;
         z \mapsto \frac{\sum m_i z_i \bar{z_i}}{\sum z_i \bar{z}_i}
            -\nu
\end{equation}
where $\nu$ is some real number. By varying $\nu$ we can obtain all
possible symplectic quotients of the action (\ref{lam}).
The special map $\mu_0$ will be denoted by $\mu$. In the sequel,
$X$ will denote the reduced phase space $\mu^{-1}(\nu)/S^1$.

As one can see from this expression the values of $\mu_\nu$ at
the fixed points equal the numbers $m_i-\nu$. Shifting of
the $m_i$ by the same number can be absorbed by a change
of $\nu$.
The requirement that $0$ is a regular value of $\mu_\nu$
translates
into the requirement that none of
the $m_i-\nu$ equals zero.

\subsection{The cohomology ring}

Recall that our task is to compute the
kernel of the ring epimorphism $\bar{r}: H_G(M) \rightarrow
H(X)$, using Poincar\'e duality on $X$ and the integration formula.
We will first determine $H_G(M)$ for our case.
\begin{prop}
    The algebra homomorphism $h:{\bf R}[\phi,\tau] \rightarrow
    {\bf R}[\phi] \otimes \Omega({\bf CP}^n)$, which sends
    $\phi$ to $\phi \otimes 1$ and $\tau$ to $1 \otimes \sigma -
    \phi \otimes \mu$ induces a surjective algebra homomorphism
    $\bar{h}:{\bf R}[\phi,\tau] \rightarrow H_{S^1}({\bf CP}^n)$.
\end{prop}

{\bf Proof}.
 From theorem \ref{thmKi} and the fact that $H({\bf CP}^n)$ is
generated by the class of the symplectic form $\sigma$,
it follows that, as a vector space,
\begin{equation}\label{33}
   H_{S^1}({\bf CP}^n) \simeq {\bf R}[\phi] \otimes
     {\bf R}[\sigma]/\sigma^{n+1}
\end{equation}
where $\phi$ and $\sigma$ both have degree two.
Since this isomorphism comes from the degeneration of a
spectral sequence, we can carry out the zig-zag construction ([BT])
and obtain that
$  H_{S^1}({\bf CP}^n) $ is generated by two elements, namely
$\phi$ and $\tau=1 \otimes \sigma - \phi \otimes \mu$.

\vspace{5pt}

{\bf Remark}

4.4) The kernel of $\bar{h}$ is equal to the ideal generated by
$\prod_{P \in F} (\tau + \mu(P) \phi)$. Thus we obtain the
following isomorphism of algebras
\begin{equation}
        H_{S^1}({\bf CP}^n) \simeq
        {\bf R}[\phi,\tau]/(\prod_{P \in F} (\tau +\mu(P) \phi))
\end{equation}
The following arguments will prove this statement. From (\ref{33})
we know $H_{S^1}({\bf CP}^n)$ as a graded vector space over ${\bf R}$.
If $l \leq n$, then dim$H_{S^1}^{2l}({\bf CP}^n)=l+1$ and it is
spanned by $\phi^{l-i} \tau^i$ ($i=0, \ldots ,l$).
If $l \geq n$, then dim$\; H_{S^1}^{2l}({\bf CP}^n) = n+1$.
So the kernel of $\bar{h}$ is a principal ideal generated by a
polynomial of degree $n+1$. To determine this polynomial note that
the restriction $H_{S^1}^{2n+2}({\bf CP}^n) \rightarrow
H_{S^1}^{2n+2}(F) \simeq {\bf R}  \phi^{n+1} \otimes {\bf R}^F$
is surjective (due to the fact that all $\mu(P_i), \; P_i \in F$
are different; see also next section), hence injective.
Furthermore, $\prod_{P \in F} (\tau + \mu(P) \phi)$ vanishes, when
restricted to $F$, from which the result follows.

\vspace{5pt}

Composing $\bar{h}$ with $\bar{r}$  we obtain a
surjective homomorphism of rings
\begin{equation}
    \bar{r} \circ \bar{h} : {\bf R}[\phi,\tau] \rightarrow H(X)
\end{equation}
Thus we see that $H(X)$ is generated by the reduced symplectic
form $\sigma_\nu$ and the curvature d$\theta$
of the circle bundle $Z \rightarrow X$.
This follows from (\ref{rmap}) and the remark
after (\ref{16}).
We shall determine the kernel $I \subset {\bf R}[\phi,\tau]$ of
this map. For sure $I$ is a non-zero ideal because $H(X)$ has no
non-zero elements of degree greater than $n-1$. In particular,
$\prod_{P \in F} (\tau +\mu(P) \phi) \in I$, which also
follows from remark 4.6. In fact, $I$ is generated by two
factors of this polynomial.
Let $k$ be the number of fixed points $P$ at which
$\mu(P)>\nu$. Define
\[     p_k= \prod_{\mu(P)>\nu}
         (\tau + \mu(P) \phi), \hspace{20pt}
       q_{n-k+1}= \prod_{\mu(P)<\nu}
           (\tau + \mu(P) \phi).
\]
The following theorem gives the cohomology ring of the
symplectic quotient $X$.
A proof of this theorem will be given later on.

\begin{thm}\label{thm5.1}
$I \subset {\bf R}[\phi,\tau]$ is the ideal
generated by $p_k$ and $q_{n-k+1}$, so
we have the following isomorphism of rings
\begin{equation}\label{cohring}
   H(X) \simeq {\bf R}[\phi,\tau]/I
\end{equation}
Explicitly, $H(X)$ is generated by the image of $\phi$,
the curvature class $c=[{\rm d}\theta]$ in $H^2(X)$ of the
circle bundle $\mu^{-1}(\nu) \rightarrow X$, and the image
of $\tau$, which is equal to $[\sigma_\nu] - \nu c$, where
$\sigma_\nu$ denotes the symplectic form of the reduced phase
space. $I$ is the ideal of relations between these generators.
\end{thm}

{\bf Remark}

4.5) The generators $\phi$ and $\tau$ map in fact to rational classes
on $X$, so ${\bf Q}[\phi, \tau] \rightarrow H(X;{\bf Q})$ is
well defined  and surjective. This follows from the following
considerations. $X$ is the quotient of a $T^2=S^1 \times S^1$ action
on the subspace $Y$ of ${\bf C}^{n+1}$ given by the equations
$\sum \bar{z}_i z_i =1$ and $\sum m_i \; \bar{z}_i z_i=\nu$.
The first $S^1$ acts linearly on ${\bf C}^{n+1}$ with all weights
equal to 1 and the second $S^1$ acts with weights $m_i$.
Let us give their generating vector fields the names
$V'$ and $V$, respectively. We will construct a connection
for this $T^2$-action such that the curvature components
on $X$ are the images of $\phi$ and $\tau$, hence rational.
Let $\theta$ be the connection of section four but now seen as
a basic element on $Y$ with
respect to the first circle action. Let $\theta '$
be the connection on $S^{2n+1}$ whose curvature gives rise to
the symplectic form on ${\bf CP}^n$, $\theta ' =
\frac{1}{2\pi} \sum {\rm Im} (\bar{z}_i {\rm d} z_i)$.
Then, on $Y$, $\iota_V \theta' = \nu$, so connection components
are $\theta$ and $\theta' - \nu \theta$ with corresponding
curvature components d$\theta$ and $\sigma - \nu {\rm d} \theta$,
precisely the differential forms to which $\phi$ and $\tau$ are
mapped.

\subsection{Bilinear forms}

Let $\alpha=\sum \alpha_j \phi^j$ and $\beta=\sum \beta_j \phi^j$
be equivariant closed forms of degrees $2l$ and $2n-2l-2$,
respectively. For convenience, we shall use the same symbols
for the cohomology classes they represent.
Let $k$ be the number of fixed points $P_i$ for
which $\mu(P_i) > \nu$. We may assume that $2k \leq n+1$, otherwise
we take the other fixed points.
Our fixed point formula gives
\begin{equation}\label{blf}
   \int_X r(\alpha) \wedge r(\beta) =
   \sum_{i=1}^k c_i \; \alpha^{(0)}(P_i) \beta^{(0)}(P_i)
\end{equation}
for certain non-zero real coefficients $c_i$. This is a bilinear
form on $H_G^{2l}(M) \times H^{2n-2l-2}_G(M)$ representing integration
on $X$.
If $\int_X r(\alpha) \wedge r(\beta)=0 $ for all $\beta \in
H_G^{2n-2l-2}$ then $\alpha$ is in the ideal $I$
(see also (\ref{pd})).
Since $I$ is non-empty, this form must be degenerated.

Note that the rhs of (\ref{blf}) can also be viewed as a
diagonal bilinear form on
${\bf R}^k \times {\bf R}^k$. As such, it is non-degenerate because
 the  $c_i$ are non-zero.
We will use this to compute $I$.

For each $l<n$ define the map
\begin{equation}
   \epsilon_l : H^{2l}_G(M) \rightarrow {\bf R}^k, \; \;
   \alpha \mapsto (\alpha^{(0)}(P_1), \ldots , \alpha^{(0)}(P_k))
\end{equation}
It is precisely the images of this map that appear on the rhs
of (\ref{blf}).
Since $H^{2l}_G(M)$ is spanned by $\phi^{l-i} \tau^{i}$ ($i=0, \ldots,
l$) as a vector space, it is easy to see
what the image of $\epsilon_l$ is. If we define $\mu_i := -\mu(P_i)$
($i=1, \ldots ,k$), then
\begin{equation}\label{vektortjes}
   \epsilon_l (H^{2l}_G(M)) =  \; < \left( \begin{array}{ccc}
                                  1\\ \vdots \\ 1 \end{array}
                            \right) ,
                            \left( \begin{array}{ccc}
                                \mu_1 \\ \vdots \\ \mu_k \end{array}
                            \right) ,
           \ldots ,
                     \left( \begin{array}{ccc}
                        \mu_1^l \\ \vdots \\ \mu_k^l \end{array}
                     \right) >
\end{equation}
Recall  that the $\mu_i$ is just
a subset of the $-m_i$
and that these are all different. Therefore, as long as
$l \leq k-1$ the image is exactly $(l+1)-$dimensional,
so $\epsilon_l$ is injective for $l<k$. Furthermore, $\epsilon_l$ is
surjective for $l \geq k-1$.

The following proposition is  now obvious.
\begin{prop}\label{52}
Let $\alpha$ be a polynomial in $\phi$ and $\tau$,
homogeneous of degree $l$, then $\alpha \in I$ iff it
is in the orthocomplement of $\epsilon_{n-l-1}
(H_G^{2n-2l-2}(M)) \subset {\bf R}^k$ for the
bilinear form diag($c_1, \ldots ,c_k$) on ${\bf R}^k$.
\end{prop}

\subsection{Computing the ideal}

In this section  we will give a proof of theorem \ref{thm5.1} by
computing the ideal $I$.
Before we compute the degeneracy of (\ref{blf})
note that if $\epsilon_l$ has a kernel then this kernel is
also in $I$. In fact, there is a kernel and we shall describe it now.
The map $\epsilon_l$ is given by
\begin{equation}
   \sum_{i=0}^l a_i  \; \phi^{l-i} \tau^i \mapsto
   ( \sum a_i \mu_1^i, \ldots , \sum a_i \mu_k^i)
\end{equation}
the $a_i$ being real coefficients. As one sees from this
expression, elements in the kernel come from polynomials
$\sum a_i x^i$ for which all the $\mu_i$ are zeroes.
Thus we have proved

\begin{prop}\label{5.1}
For $l<k$ the maps $\epsilon_l$ are injective. For
$l \geq k-1$ they are surjective and if $l \geq k$
the kernel of $\epsilon_l$ consists of all polynomials in
$\phi$ and $\tau$, homogeneous of degree $l$, who are
divisible by $p_k$, where $p_k$ is the polynomial
$\prod_j (\tau - \mu_j  \phi)$ of degree $k$.
\end{prop}

This implies that the ideal generated by this polynomial
is a subideal of $I$. Moreover, it expresses $\tau^k$ in
lower order powers of $\tau$ and therefore gives already a
bound, equal to $k$, on the Betti numbers of $X$.

To obtain the other polynomials in $I$ we proceed as follows.
For each $l$, we know the image of $\epsilon_{n-l-1}$.
If $n-l-1 \geq k-1$ then $\epsilon_{n-l-1}$ is surjective,
so from proposition \ref{52} we derive that there are no
$\alpha \in I$ of degree $l \leq n-k$.
Since $k \leq \frac{n+1}{2}$, $n-k \geq \frac{n-1}{2}$, which is
half of the dimension of $X$. This implies that all
polynomials in $I$ have at least degree $k$ and therefore
$b_l(X)=l+1$ for all $l<k$. Furthermore, for $k \leq l \leq n-k$,
the only relations are those coming from ker($\epsilon_k$). So,
for these $l$, $b_l(X)=k$. For $l>n-k$, the $b_l(X)$ equal
the $b_{n-l-1}(X)$ due to Poincar\'e duality. Thus, they
decrease by one if $l$ increases by one.

The next proposition finishes the proof of theorem \ref{thm5.1}.
\begin{prop}
Let $\nu_i$ denote the $n-k+1$ values $\mu(P_i)$, where $P_i$ are
critical points at which $\mu(P_i) < \nu$.
Then the polynomial
$q_{n-k+1} = \prod_j (\tau + \nu_j \phi)$ is in I.
Moreover, the ideal I is generated by $p_k$ and $q_{n-k+1}$.
\end{prop}

{\bf Proof}.
If we had taken a summation over the fixed points $P_i$ for
which $\mu(P_i)<\nu$, then we would have found $q_{n-k+1}$
instead of $p_k$ to be in $I$. This proves that it is in $I$.
Furthermore, the two polynomials have no common factors, as all
the $\mu(P)$, $P \in F$,
are different. This implies that the
subideal generated by these two polynomials gives mutually dependent
relations on the generators
only in degree greater than $(n-k+1) + k=n+1$.
So there is no need for more generators.

\vspace{10pt}

{\bf Remarks}

4.6) For non-isolated fixed points it happens that the cohomology ring
is given by the same formula (\ref{cohring}). E.g., if $l$ of the $m_i$
are equal, there is a ${\bf CP}^{l-1}$ fixed point manifold. The
momentum map $\mu_\nu$ is constant on this submanifold and instead of
$l$ different $\mu(P_i)$ we have to take $l$ times the same value,
so that the total number of $\mu_i$ and $\nu_i$ remains $n+1$.

However, the proof in this case is somewhat different. First of all
the formula (\ref{blf}) involves higher cohomology classes, namely
all possible classes on the fixed point manifolds. This implies that
the associated bilinear form, using (\ref{genfpf})
rather than (\ref{blf}),
is no longer diagonal but consists
of triangular blocks of dimension equal to the cohomological
dimension of
the fixed point manifold (in this case its complex dimension
plus one). Furthermore, in the maps $\epsilon_i$
there will appear not only constants, but also cohomology classes.
With these changes the proof goes entirely through for the
general case.

4.7) The symplectic quotients studied here are
toric varieties. The cohomology rings of these
spaces have been computed
by Danilov ([Da]). This provides a different description, obtained
by different methods, of the same cohomology rings.

4.8) Very recently, Lisa Jeffrey and Frances Kirwan came up with
a fixed point formula for general Lie groups ([JK]). Although
inspired by Witten ([W3]), they use tough Fourier analysis to
prove their formula.

\section*{acknowledgements}
I am grateful to a lot of people for discussions and helpful
comments. Especially, I like to thank Hans Duistermaat and
Peter Braam for their great help the last four years.
Furthermore, I like to thank Raymond Stora, Gijs Tuynman,
Jim Stasheff and Takashi Kimura for useful visits to
their departments and Sophie Chemla for the work we did
together.

\chapter*{References}

\begin{description}
\item[[AB]] M. Atiyah, R. Bott, The moment map and equivariant
                 cohomology,
                 Topology 23 (1984) 1.
\item[[AJ]] M. Atiyah, L.C. Jeffrey, Topological Lagrangians and
               cohomology, J. of Geom. and Phys., vol.7,
                no1 (1991) 119
\item[[AM]] R. Abraham, J.E. Marsden, Foundations of Mechanics,
                Addison Wesley, New York 1987
\item[[BGV]] N. Berline, E. Getzler, M. Vergne, Heat Kernels and
                Dirac Operators, Springer Verlag 1991
\item[[BRS]] C. Becchi, A. Rouet, R. Stora, Renormalization
                of the abelian Higgs-Kibble model, Commun. Math. Phys.
                42 (1975) 127
\item[[BS]] L. Baulieu, I. Singer, Topological Yang-Mills symmetry,
                 Nucl. Phys. (proc. suppl.) 5B (1988) 223
\item[[BT]] R. Bott, L. Tu, Differential Forms in Algebraic Topology,
                     GTM 82, Springer Verlag 1982
\item[[BV]] I.A. Batalin, G.A. Vilkovisky, Relativistic S-matrix
                 of dynamical systems with boson and fermion
                 constraints, Phys. Lett. 69B (1977) 309
\item[[Ca]] H.Cartan, Transgression dans un groupe de Lie et dans
                un espace fibr\'e principal, Colloque de Topologie CBRM
                  Bruxelles (1950) 57-71
\item[[CK]] S. Chemla, J. Kalkman, BRST cohomology for certain
                 reducible topological symmetries, to appear in
                 Commun. Math. Phys.
\item[[Da]] V.I. Danilov, The geometry of toric varieties,
                 Russ. Math. Surveys 33 (1978) 97
\item[[DH]] J.J. Duistermaat, G.J. Heckman, On the variation in the
                 cohomology of the symplectic form of the reduced phase
                 space, Invent. Math. 69 (1982) 259;
                 Addendum, Inv. Math. 72 (1983) 153
\item[[FF]] E.S. Fradkin, T.E. Fradkina, Quantization of relativistic
                 systems with boson and fermion first and
                 second class constraints, Phys. Lett. 72B (1978) 343
\item[[FHST]] J. Fisch, M. Henneaux, J. Stasheff, C. Teitelboim,
                 Existence, uniqueness and cohomology of the classical
                 BRST charge with ghosts of ghosts, Commun.
                 Math. Phys. 120 (1989) 379
\item[[Gi]] V.A. Ginzburg, Equivariant cohomologies and
                 K\"ahler geometry,
                 Funct. Anal. and its Appl. 21 (1987) 271
\item[[He]] M. Henneaux, Hamiltonian form of the path integral
                 for theories with a gauge freedom,
                 Phys. Rep. 126 (1985), no1
\item[[HT]] M. Henneaux, C. Teitelboim, BRST cohomology in classical
                 mechanics, Commun. Math. Phys. 115 (1988) 213
\item[[Hu]] D. Husemoller, Fibre Bundles, GTM 20, Springer Verlag
\item[[JK]] L.C. Jeffrey, F.C. Kirwan, Localization for non-abelian
                 group actions, alg-geom/9307001
\item[[Ka1]] J. Kalkman, BRST model for equivariant cohomology and
                 representatives for the equivariant Thom class,
                 Commun. Math. Phys. 153 (1993) 447
\item[[Ka2]] J. Kalkman, Cohomology rings of symplectic quotients,
                 preprint 795, Math. Institute, University of Utrecht
\item[[Ki]] F.C. Kirwan, Cohomology of Quotients in Symplectic
                   and Algebraic Geometry, Math. Notes Vol. 31,
                   Princeton University Press 1984
\item[[KS]] B. Kostant, S. Sternberg, Symplectic reduction,
                 BRS cohomology and infinite dimensional Clifford
                 algebras, Ann. of Phys. 176 (1987) 49
\item[[MQ]] V. Mathai, D. Quillen,
                 Thom classes, superconnections
                   and equivariant differential forms,
                 Topology 25 (1986) 85.
\item[[OSvB]] S. Ouvry, R. Stora and P. van Baal, Algebraic
                 characterization of TYM, Phys. Lett. 220B (1989) 1590
\item[[Sa]] I. Satake, On the generalization of the notion of
                 manifold, Proc. Nat. Acad. Sci. USA 42 (1956) 359
\item[[St]] J.D. Stasheff, Homological (ghost) approach to
                 constrained Hamiltonian systems, Contemp. Math.
                 132 (1992) 595
\item[[W1]] E. Witten, Topological quantum field theory, Commun. Math.
                 Phys. 117 (1988) 353
\item[[W2]] E. Witten, Introduction to cohomological field theories,
                 Int. J. Mod. Phys. A6 (1991) 2775
\item[[W3]] E. Witten, Two dimensional gauge theories revisited,
                 J. Geom. Phys. 9 (1992) 303
 \item[[dW]] B. de Witt, Supermanifolds, Cambridge University Pres,
                         Second Edition, 1992.
\item[[Wu]] S. Wu, An integration formula for the square of moment
                 maps of circle actions, hep-th/9212071

\end{description}

\end{document}